\documentclass[12pt,fleqn]{article}

\usepackage{amsmath} 

\renewcommand{\a}{\alpha} 
\renewcommand{\b}{\beta}
\newcommand{\g}{\gamma} 

\renewcommand{\d}{\delta} 
\newcommand{\D}{\Delta}
\renewcommand{\k}{\kappa}

\newcommand{\m}{\mu} 
\newcommand{\n}{\nu}
\newcommand{\p}{\pi} 
\renewcommand{\r}{\rho} 
\newcommand{\s}{\sigma}

\renewcommand{\t}{\tau}

\newcommand{\hx}{{\hat{x}}} 
\newcommand{\hbx}{{\hat{\bf x}}} 
\newcommand{\hbX}{{\hat{\bf X}}} 
\newcommand{\hX}{{\hat{X}}} 

\newcommand{\tl}{\lambda}

\newcommand{\cF}{{\cal F}} 
\newcommand{\cL}{{\cal L}}

\newcommand{\sfrac}[2]{{\textstyle\frac{#1}{#2}}}

\newcommand{\pd}{\partial}
\newcommand{\pda}[2]{{\frac{\pd #1}{\pd #2}}}
\newcommand{\pdb}[3]{{\frac{\pd^2 #1}{\pd #2\pd #3}}}
\newcommand{\pdc}[2]{{\frac{\pd^2 #1}{\pd^2 #2}}}
\newcommand{\spda}[2]{{\sfrac{\pd #1}{\pd #2}}}

\newcommand{\nn}{\nonumber}

\newcommand{\half}{\frac{1}{2}}
\newcommand{\shalf}{{\textstyle\frac{1}{2}}}

\newcommand{\bal}{{\mbox{\boldmath $\alpha$}}}
\newcommand{\bbe}{{\mbox{\boldmath $\beta$}}}
\newcommand{\bth}{{\mbox{\boldmath $\theta$}}} 
\newcommand{\brh}{{\mbox{\boldmath $\rho$}}}
\newcommand{\bn}{{\bf n}}

\newcommand{\bx}{{\bf x}}
\newcommand{\bX}{{\bf X}}
\newcommand{\by}{{\bf y}}
\newcommand{\bz}{{\bf z}}
\newcommand{\bR}{{\bf R}}

\newcommand{\isum}{\sum_{i=0}^N}
\newcommand{\rsum}{\sum_{r=1}^R}
\newcommand{\msum}{\sum_{m=1}^M}
\newcommand{\nsum}{\sum_{{\bf n}\geq 0}}

\begin{document}

\title{Symmetries in jump-diffusion models with
applications in option pricing and credit risk}
\author{J.K. Hoogland \& C.D.D. Neumann\\[1ex]
  {\scriptsize CWI, P.O. Box 94079} \\[-.5ex]
  {\scriptsize 1090 GB Amsterdam  } \\[-.5ex]
  {\scriptsize The Netherlands } \\[-.5ex]
  {\scriptsize Email:
    {\tt $\{$jiri.hoogland,dimitri.neumann$\}$@cwi.nl} } \\[2ex]
    M.H. Vellekoop \\[1ex]
  {\scriptsize Department of Applied Mathematics} \\[-.5ex]
  {\scriptsize University of Twente} \\[-.5ex]
  {\scriptsize P.O. Box 217, 7500 AE, Enschede} \\[-.5ex]
  {\scriptsize The Netherlands} \\[-.5ex]
  {\scriptsize Email: {\tt m.h.vellekoop@math.utwente.nl}} }
\date{First version: October 25, 1917\\
      Current version: \today}
\bibliographystyle{plain}

\maketitle
\thispagestyle{empty}

\begin{abstract}
It is a well known fact that local scale invariance plays a
fundamental role in the theory of derivative pricing. Specific
applications of this principle have been used quite often under
the name of `change of numeraire', but in recent work it was shown
that when invoked as a fundamental first principle, it provides
a powerful alternative method for the derivation of prices
and hedges of derivative securities, when prices of the 
underlying tradables are driven by Wiener processes. In this
article we extend this work to the pricing problem in markets
driven not only by Wiener processes but also by Poisson processes,
i.e. jump-diffusion models. It is shown that in this case too,
the focus on symmetry aspects of the problem leads to important
simplifications of, and a deeper insight into the problem.
Among the applications of the theory we consider the pricing of
stock options in the presence of jumps, and Levy-processes. Next
we show how the same theory, by restricting the number of jumps,
can be used to model credit risk, leading to a `market model' of
credit risk. Both the traditional Duffie-Singleton and
Jarrow-Turnbull models can be described within this framework, 
but also more general models, which incorporate default
correlation in a consistent way. As an application of this theory
we look at the pricing of a credit default swap (CDS) and a
first-to-default basket option.
\end{abstract}

\section{Introduction}
\label{sec:introduction}

In the last thirty years, a considerable amount of scientific effort
has been put into attempts to generalize the highly successful
derivative pricing methods of Black and Scholes \cite{BlackScholes73}
to so called jump-diffusion models. In these models, the Wiener
processes which were used by Black and Scholes to model
stock dynamics, are complemented by Poisson processes which cause
unprevisible discontinuities in the stock price \cite{Merton76}.
Such discontinuities are needed to incorporate unexpected sudden
market events in price models. Possible applications are the modeling
of stock market crashes, interest rates and credit risk. In general,
the addition of Poisson processes causes the market model to be
incomplete, since the extra stochastic factors cannot be offset by
existing contracts, and this makes perfect replication of derivatives
impossible.

The traditional way to circumvent this problem is to use standard
no-arbitrage pricing theory, in which prices are calculated as an
expectation under some equivalent martingale measure. It is well known
that for an arbitrage free but incomplete market such a measure exists
but is not unique \cite{HarrisonPliska81}.

As an example consider the class of intensity based (or reduced form)
credit risk models, such as Jarrow-Turnbull \cite{JarrowTurnbull95} and
Duffie-Singleton \cite{DuffieSingleton95}. In these models a martingale
measure is postulated using an implied risk free default intensity,
which is then fitted to market prices of e.g. corporate bonds. This
approach has some marked disadvantages. For example, the relation
between the real world measure and the equivalent martingale measure
is not clear. Furthermore, it is not obvious what hedge ratios should
be used (certainly not the usual Black Scholes deltas).

An alternative approach to the problem is to assume that the
market, if not already complete, can be made complete by the
introduction of additional contracts. The non-uniqueness in the
pricing problem then translates into the freedom that one has in
the specification of these new contracts. After completion of the
market, prices and hedging strategies can be determined uniquely
\cite{BeumeeHilberink99, VellekoopBeumeeHilberink99,JarrowMadan95,
Jensen99}. The question remains under what circumstances a market
{\it can} be made complete. In Ref.~\cite{VellekoopBeumeeHilberink99},
for example, it was shown that if, for a given asset, the stochastic
process which causes a jump is the only factor in the dynamics which
is idiosyncratic and the percentage of loss upon a jump in the price
is fixed and known a priori, it is possible to complete the market
by the introduction of an {\sl insurance contract} which pays a
certain amount of money once a jump has occurred in exchange for
an insurance premium during the lifetime of the contract.

In this article, we follow the alternative `complete market'
approach, making use of the {\sl tradable formalism} as introduced
in Refs. \cite{HooglandNeumann99a, HooglandNeumann99b}, thereby
extending the powerful symmetry techniques to the realm of
jump-diffusion models. We start out with a model which is
complete. The core of the formalism is the idea that pricing problems
should be formulated only in terms of self-financing objects, which we
call tradables. From simple dimensional analysis one then finds that
derivative prices must be homogeneous functions of degree one in these
tradables. The consequent use of these proper coordinates makes
numeraire invariance manifest, and the pricing problem more
transparent. We find
\begin{itemize}
\item Pricing equations in the form of partial differential difference
  equations (PDDE's) for which the numeraire invariance of their solutions
  is manifest
\item Explicit expressions for hedge ratios
\item Analytic solutions for European style contracts in a certain
  class of jump-diffusion models
\end{itemize}
The relation between our approach and the martingale approach is
illustrated, leading to a direct relation between real world and
martingale measures. It is shown that the usual notion of a 'market
price of risk' is not a numeraire invariant quantity for Poisson
processes, in contrast to the case of Wiener processes. Therefore it
is only well-defined with respect to an arbitrary choice of a tradable
which is declared to be 'risk free' (usually the treasury bond).

As was mentioned before, the model, although derived under the
assumption of market completeness, can deal with incomplete markets
too, in as far as these markets can be {\it made} complete by the
introduction of additional contracts for each independent source
of jump risk. The freedom that we have in the specification of these
contracts allows to introduce a specific market price of jump risk in
the model. We illustrate this by considering the pricing of a derivative
depending on a stock and a bond, where the stock is driven by
one Wiener and several Poisson processes, taking the bond as
numeraire. In a certain limit, this indicates how to price such
contracts if the stock is driven by a Levy-process.

By restricting the number of possible jumps of the individual
Poisson processes to one, the very same theory can be applied to
credit risk modeling. This approach could be called a `market model'
approach to credit risk, in the sense that we directly model the real
world dynamics of the underlyings. Although conceptually different,
the approach is closest to the intensity based approach and in fact
we show that it is able to unify the models of Jarrow-Turnbull and
Duffie-Singleton in one framework. It also allows the incorporation
of default correlation in a straightforward and consistent way.

The outline of this article is as follows. In section 2 we refresh
the basic ideas behind the tradable formalism. In section 3 we
rederive pricing equations for pure diffusion models. Section 4 is
the theoretic core of the article. It contains a general theory for
pricing in jump-diffusion models, and investigates its symmetry
properties. Some applications are given. In section 5, we look at
consequences for pricing if Poisson processes are restricted to one
jump only, having credit risk in mind. We review the Duffie-Singleton
and Jarrow-Turnbull approaches and show that they both fit in the
model and lead to almost identical pricing equations. As an
application we calculate the value of a credit default swap in the
two models. Next we look at multiple asset credit risk models,
firstly considering the pricing of a first-to-default basket option,
secondly deriving a consistent model with default correlation.

\section{Exploiting symmetries}
\label{sec:expl-symm}

Any system of prices should be invariant under
appropriate scaling transformations. Indeed, since the value of assets
can only be defined in terms of the values of other assets, an
economy-wide uniform proportional change of the values of all possible
assets would simply amount to the introduction of a new measurement
unit for value, and leave the economy essentially unchanged. Specific
applications of this principle have been used quite often in finance
under the name of 'change of numeraire', but in the recent article
\cite{HooglandNeumann99a} it was shown that when invoked as an overall
fundamental first principle, it provides a powerful alternative
method for the derivation of prices and hedges for derivative securities
on tradable assets.

We assume a market in which $N+1$ tradables exist, and we denote the
price at time $t$ of asset $i$ by $X^i_t$ with $i=0,\ldots,N$ and
$\bX_t\equiv (X^0_t,\ldots,X^N_t)$. These prices (in whatever a priori
fixed numeraire we would like to choose) are modeled as stochastic
processes $\{ X^i_t,t\geq 0\}$ on a filtered probability space
$(\Omega,\cF ,\{\cF_t\}_{t\geq 0},P)$ where the filtration satisfies
the usual assumptions. On this probability space we define canonical
(possibly multi-dimensional) Wiener and Poisson Processes in the usual
manner, which together generate the filtration $\{\cF_t\}_{t\geq 0}$.
The processes $\{ X^i_t,t\geq 0\}$ will be adapted to this filtration.

In this market we now consider the price of a European style
derivative security $V$ which pays an amount $V(\bx,T)$ at time of
maturity $T$ only depending on the value of the price vector ${\bf
x}\equiv(x_0,x_1,\ldots,x_N)$ of the tradables at maturity ${\bf
X}_T$. We assume that the market is complete, i.e. the price $V$ 
of the security at an earlier time $t\leq T$ exists and
is uniquely determined by the market, and furthermore that this
price is a function of the prices $X^i_t$ at time $t$ only.
Under these assumptions we may write the price of the derivative
security as $V(\bx,t)$ and then a dimensional analysis argument
shows that, since the prices of tradables and the security $V$
must be expressed in the same measurement units, we must have
\[
V(\bX_t,t) = \isum \phi_i(\bX_t,t)\, X^i_t
\]
for certain dimensionless quantities $\phi_i(\bX_t,t)$, i.e.
homogeneous functions of degree zero in $\bX_t$. Consequently,
the price function is homogeneous of degree $1$
\begin{equation}
V(\a\bx,t) = \a V(\bx,t)
\label{eq:1}
\end{equation}
Differentiating with respect to $\a$ and setting $\a$ equal to
one leads to the Euler formula
\begin{equation}
V(\bx,t) = \isum\pda{V(\bx,t)}{x_i}\, x_i
\label{eq:2}
\end{equation}
This simple formula is at the core of Black-Scholes pricing.
Indeed, when prices are driven only by Wiener processes, we
can prove that
\[
dV(\bX_t,t) = \isum\pda{V(\bx,t)}{x_i} \, dX^i_t
\]
given that $V$ satisfies a PDE (see next section). This shows
that a portfolio consisting of an amount
$\phi_i(\bx,t)=\smash{\pda{V(\bx,t)}{x_i}}$ in each tradable $i$
not only {\sl replicates} the security $V$, but is also 
{\sl self-financing}, i.e. all changes in the price of the portfolio
are caused by changes in the prices of the underlying tradables
themselves. No money is put in or extracted from the portfolio.

When prices are driven by both Wiener and Poisson processes,
the hedge ratios $\phi_i(\bx,t)$ turn out to be more complicated,
in general containing not only ordinary but also discrete derivatives
of the price, and a little more work is required to find them.
Nevertheless, the scale invariance expressed by Eqs. (\ref{eq:1})
and (\ref{eq:2}) still plays a fundamental role, as will be seen
repeatedly in section 4.

\section{Black-Scholes dynamics}
\label{sec:black-schol-dynam}

To illustrate the ideas outlined in the previous section,
consider the case of $N+1$ tradables with price processes $X^i_t$,
($i=0,\ldots,N$) driven by $N$ Wiener processes
\begin{equation}
\label{eq:3}
\frac{dX^i_t}{X^i_t} 
= \m_i(\bX_t,t)\, dt 
+ \sum_{k=1}^N\s_i^k(\bX_t,t)\, dW^k_t
\end{equation}
Here $\{ (W_t,\cF_t),\, t\geq 0\}$ is a standard $N$-dimensional
Wiener process. The vector functions
$\m_i:\bR^{N+1}\times\bR\to\bR$ and $\s_i:\bR^{N+1}\times\bR\to\bR$
are assumed to satisfy the standard growth conditions which guarantee
uniqueness and existence of the solutions of this stochastic
differential equation. Notice that the functions $\m_i$ and $\s_i$
should all be homogeneous functions of degree $0$. In other words,
they can be functions of ratios $X^i_t/X^j_t$ of tradable prices
only, since the substituted values have to be dimensionless.  For
these dynamics we have by It\^o's rule that
\[ 
dV(\bX_t,t) 
= 
\cL V(\bX_t,t)\, dt 
+ \isum\pda{V(\bX_t,t)}{x_i}\, dX^i_t
\] 
with
\begin{equation}
\label{eq:4}
\cL V(\bx,t) 
\equiv 
\pda{V(\bx,t)}{t} 
+ \shalf\sum_{i,j=0}^N\sum_{k=1}^N \s_i^k(\bx,t)\s_j^k(\bx,t)\, x_ix_j\,
\pdb{V(\bx,t)}{x_i}{x_j}
\end{equation}
and solving the partial differential equation $\cL V(\bx,t)=0$,
subject to contract specific boundary conditions, then leads to a
self-financing portfolio in terms of tradables, as we have seen in
the previous section\footnote{Assuming that second order partial
derivatives of the price function with respect to the tradables,
and the first order derivative with respect to time exist.}.
This leads to a unique price for European style
derivative securities before maturity $T$ by specifying the payoff
$V(\bx,T)$ as a boundary condition. Of course, other types of
contracts can be priced too, either by modifying the boundary
conditions (e.g. American type contracts) or, in the case of strong
path-dependence, by introducing specific derived tradables, which make
the problem Markovian again (e.g. Asians, see \cite{HooglandNeumann00a}).
Note that the pricing PDE does not contain drift terms. This has
important advantageous consequences when implementing numerical
schemes to solve the equation, see Ref.~\cite{HooglandNeumann00b}.

\subsection{Symmetries of the PDE}
\label{sec:symmetries-pde}

The invariance of the price of a derivative security under a change of
numeraire should be inherited by the pricing PDE. Of course, prices 
under a new numeraire can in general be expressed in terms of prices
under the old numeraire as $\hbX_t\equiv \bX_t/Y_t$, where the process
$Y_t$ which defines the numeraire change can be arbitrary and
stochastic, as long as it is positive definite. Now suppose that
$Y_t$ satisfies
\[
\frac{dY_t}{Y_t} = \n(\bX_t,t) dt + \sum_{k=1}^N g_k(\bX_t,t) dW^k_t
\]
Consistency requires that $\n$ and $g_k$ are homogeneous functions
of degree $0$ in $\bX_t$, so that $\n(\bX_t,t)=\n(\hbX_t,t)$
and similar for $g_k$. These functions should also satisfy the
standard growth conditions. By using It\^o's rule, we find that the
prices under the new numeraire satisfy
\begin{eqnarray*}
\frac{d\hX^i_t}{\hX^i_t} &=& 
\big(\m_i(\hbX_t,t)-\n(\hbX_t,t) -g_k(\hbX_t,t)(\s_i^k(\hbX_t,t)
-g_k(\hbX_t,t))\big)\, dt \\
&&+ \sum_{k=1}^N\big(\s_i^k(\hbX_t,t)-g_k(\hbX_t,t)\big)\, dW^k_t
\end{eqnarray*}
If we write down the pricing PDE in terms of the rescaled
tradables, we find extra terms, proportional to the $g_k(\hbx,t)$:
\begin{eqnarray*}
\lefteqn{\cL V(\hbx,t) 
= \pda{V(\hbx,t)}{t} }\\
&&+ \shalf\sum_{i,j=0}^N\sum_{k=1}^N
\big(\s_i^k(\hbx,t)-g_k(\hbx,t)\big)\big(\s_j^k(\hbx,t)-g_k(\hbx,t)\big)\, 
\hx_i\hx_j\, \pdb{V(\hbx,t)}{\hx_i}{\hx_j} = 0
\end{eqnarray*}
Now it is exactly the homogeneity property of $V$ which ensures
that these new terms vanish. Indeed these terms can be written as
\[
\shalf \sum_{k=1}^N \isum g_k(\hbx,t)
(g_k(\hbx,t)-2\s_i^k(\hbx,t))\,\hx_i
\left( \sum_{j=0}^N\hx_j\, \pdb{V(\hbx,t)}{\hx_i}{\hx_j}\right)
\]
and, by differentiating Eq.~(\ref{eq:2}) once more,
we find that for all $i$
\[
\sum_{j=0}^N \hx_j\pdb{V(\hbx,t)}{\hx_i}{\hx_j}=0
\]
We see that the solutions of the PDE for the price of a derivative
security are invariant under changes of the numeraire. Furthermore,
the price function itself should be invariant under the substitutions
\[
\s^k_i(\bx,t)\to\s^k_i(\bx,t)-g_k(\bx,t)
\hspace{5mm} \mbox{for all $i$}
\]
A clever choice of numeraire can simplify the computation of
derivative security prices significantly. Indeed, picking tradable
$0$ as numeraire is easily seen to be equivalent to the choice
$g_k=\s^k_0$ and $\n=\m_0$. The price process of this tradable
then becomes trivial $d\hbX_t=0$ and the dimension of the PDE
reduces by one. This technique was used in
Refs.~\cite{HooglandNeumann00a, HooglandNeumann00c} to derive compact
PDE's for the pricing of arithmetic Asian and passport options.

How does this all relate to the usual BS-framework? In that case,
the interest rate is constant $r$ and there is a fixed numeraire,
namely money. One of the tradables, say $\bX^0$, will be the riskless
bond, satisfying $d\bX^0=r\bX^0dt$, and so $\bX^0\sim e^{rt}$
in units of money. Now since this is a deterministic function,
it can be eliminated from the PDE, Eq.~(\ref{eq:4}), leading to
the familiar Black Scholes PDE. The disadvantages are obvious:
this PDE contains drift terms, which can lead to numerical problems
when numerically solving it and the manifest numeraire invariance
is lost. In contrast, in our formalism the interest rate would
only appear indirectly in prices via the value of $\bX^0$.

\subsection{A simple example}
\label{sec:simple-example-1}

Consider two tradables with price processes $X^{0,1}_t$ driven by
one Wiener process $W_t$:
\[
\frac{dX^i_t}{X^i_t} = \m_i(X^0_t,X^1_t,t)\, dt +
\s_i(X^0_t,X^1_t,t)\, dW_t, \hspace{5mm} (i=0,1)
\]
A European security with payoff $f(x_0,x_1)$ at maturity $T$ will
have value $V(x_0,x_1,t)$ at time $t$, and this value should satisfy
the following PDE
\[
\pda{V(x_0,x_1,t)}{t} 
+ \shalf\sum_{i,j=0,1}
\s_i(x_0,x_1,t)\s_j(x_0,x_1,t)\, x_ix_j\, \pdb{V(x_0,x_1,t)}{x_i}{x_j}
= 0
\]
To reduce the dimension of the PDE we pick tradable $0$ as numeraire.
Introducing $\t\equiv T-t$, $y\equiv x_1/x_0$,
$V(x_0,x_1,t)\equiv x_0v(y,\t)$, the PDE simplifies to
\[
-\pda{v(y,\t)}{\t} + \shalf \s(y,\t)^2\, y^2\, \pdc{v(y,\t)}{y} = 0
\]
where we define
\[
\s(y,\t) \equiv \s_1(x_0,x_1,T-\t)-\s_0(x_0,x_1,T-\t)
\]
Here we use the homogeneity of degree zero of the volatility
functions $\s_i$, which implies that these functions can only
depend upon the ratio $x_1/x_0$, i.e. on $y$. The boundary condition
becomes $v(y,0)=f(1,y)$.

\section{Defaultable dynamics}
\label{sec:defaultable-dynamics}

In this section we extend the pure diffusion case by adding
jump-processes, where jumpsizes of the tradables are functions
of the tradables and time.

\subsection{A simple example}
\label{sec:simple-example}

Before treating the general case we give a simple example. Consider
two tradables with price processes satisfying
\[
\label{eq:5}
\frac{dX^i_t}{X^i_{t_-}} = \m_i\, dt + (\a_i-1)\, dN_t,
\hspace{5mm} (i=0,1)
\]
Here $N_t$ denotes a Poisson process with a strictly positive
intensity, driving both price processes, and the $\m_i$ and $\a_i$
are constants for $i=0,1$. Since we want our prices to remain
strictly positive, the $\a_i$ should also be strictly positive.
No-arbitrage imposes further restrictions on the parameters:
in order that $X^1_t/X^0_t$ is not a strictly in- or decreasing
process, we must have the inequality $(\m_1-\m_0)/(\a_1-\a_0)<0$. 
We will come back to this point in Section~\ref{sec:solving-pdde}.
A derivative security price $V(\bX_t,t)$, where $\bX_t\equiv
(X^0_t,X^1_t)$, having second order derivatives w.r.t. $X^0_t$ and
$X^1_t$, and first order derivative w.r.t. $t$, then satisfies (see
Section~\ref{sec:change-variables})
\begin{eqnarray}
  dV(\bX_t,t) 
  &=& 
  \bigg(
  \pda{V(\bX_{t_-},t) }{t}
  + \sum_{i=0,1}\m_i X^i_{t_-}\pda{V(\bX_{t_-},t)}{x_i}
  \bigg)\, dt
  \nn
  \\
  &&
  +\big(
  V(\bal \bX_{t_-},t)-V(\bX_{t_-},t)
  \big)\, dN_t
  \label{eq:6}
\end{eqnarray}
Here we introduced the shorthand notation $\bal\bX_t\equiv
(\a_0X^0_t,\a_1X^1_t)$. We can rewrite Eq.~(\ref{eq:6}) now in such a
way that we get terms proportional to the $dX^i_t$ and a remainder
\begin{eqnarray*}
dV(\bX_t,t) 
&=&
\sum_{i=0,1}\phi_i(\bX_{t_-},t) \, dX^i_t
\\
&&+ \bigg(
\pda{V(\bX_{t_-},t)}{t}
+ \sum_{i=0,1} \m_i X^i_{t_-}\bigg(\pda{V(\bX_{t_-},t)}{x_i}-\phi_i(\bX_{t_-},t)\bigg)
\bigg)\, dt
\end{eqnarray*}
where we define
\begin{eqnarray}
  \phi_0(\bx,t)
  &\equiv& 
  \frac{V(\a_1 x_0,\a_1 x_1,t)-V(\a_0 x_0,\a_1 x_1,t)}{x_0(\a_1-\a_0)}
  \label{eq:7}
  \\
  \phi_1(\bx,t)
  &\equiv& 
  \frac{V(\a_0 x_0,\a_0 x_1,t)-V(\a_0 x_0,\a_1 x_1,t)}{x_1(\a_0-\a_1)}
  \label{eq:8}
\end{eqnarray}
such that
\begin{equation}
  \label{eq:9}
  x_0\phi_0(\bx,t)+x_1\phi_1(\bx,t) = V(\bx,t)
\end{equation}
Here we use homogeneity to pull the $\a_1$ and $\a_0$ out of the $V$'s
in Eqs.~(\ref{eq:7}) and ~(\ref{eq:8}) respectively. We see that the
portfolio $V(x,t)$ is self-financing if the following partial differential
difference equation (PDDE) holds.
\begin{equation}
  \label{eq:10}
  \cL V(\bx,t) \equiv \pda{V(\bx,t)}{t}
  + \sum_{i=0,1} \m_i x_i\bigg(\pda{V(\bx,t)}{x_i}-\phi_i(\bx,t)\bigg)
  = 0
\end{equation}
Indeed, in that case we have the self-financing condition,
by Eq.~(\ref{eq:9})
\[
  dV(\bX_t,t) = \sum_{i=0,1}\phi_i(\bX_{t_-},t)dX^i_t
\]
and we see that the $\phi_i(\bx,t)$ are hedge ratios for
the contract.

\subsubsection{Symmetries of the PDDE}
\label{sec:symmetries-pde-1}

Again, the PDDE should be invariant under a change of numeraire.
So let us write $\hbX_t\equiv\bX_t/Y_t$ for prices under a new numeraire,
where the process which defines the numeraire change satisfies
\[
\frac{dY_t}{Y_{t_-}} = \k\, dt + \big(\g-1\big)\, dN_t
\]
The choice of $\k$ and $\gamma$ is arbitrary with the constraint
that $\g>0$, since numeraires should be positive definite. Then
the price processes of the two tradables, expressed in terms of
the new numeraire, become
\begin{equation}
  \label{eq:11}
  \frac{d\hX^i_t}{\hX^i_{t_-}} = (\m_i-\k)\, dt +
  \bigg(\frac{\a_i}{\g}-1\bigg)\, dN_t
\end{equation}
Remember that homogeneity ensures that the functional form of $V$
is invariant under different numeraires, thus
$V(X^0_t,X^1_t,t)\equiv Y\,V(\hX^0_t,\hX^1_t,t)$. How does the change
of numeraire effect the hedge ratios, Eqs.~(\ref{eq:7}) and
(\ref{eq:8})? The hedges are homogeneous of degree $0$ in the tradables.
In other words, they only depend on the ratio $\a_1/\a_0$ and
therefore they are invariant under changes of numeraire,
$\phi_i(\bx,t)=\phi_i(\hbx,t)$, as they should be. Now consider the
PDDE, derived under the new numeraire
\[
\cL V(\hbx,t) = \pda{V(\hbx,t)}{t} + \sum_{i=0,1}(\m_i-\k) \hx_i
\bigg(\pda{V(\hbx,t)}{\hx_i}-\phi_i(\hbx,t) \bigg) = 0
\]
We see that it picks up extra terms, proportional to $\k$,
\[
\sum_{i=0,1}\hx_i\bigg(\pda{V(\hbx,t)}{\hx_i}-\phi_i(\hbx,t)\bigg)
\]
But these vanish because of homogeneity together with the
replication condition Eq.~(\ref{eq:9}) and so the PDDE is indeed
invariant under changes of numeraire. Just as we could write down a
PDE for the price of a derivative security in the pure diffusion case,
it is possible to write down a PDDE for the pure jump case with fixed
jump size, which is invariant under changes of numeraire. If we solve
this PDDE, we find a price for the derivative security, depending on
the two tradables with prices $X^0_t$ and $X^1_t$, together with the
hedge ratios Eqs.~(\ref{eq:7}) and (\ref{eq:8}).

\subsubsection{Solving the PDDE}
\label{sec:solving-pdde}

To solve Eq.~(\ref{eq:10}) we introduce the notation $\t\equiv T-t$,
$y_i\equiv e^{-\m_i\t}x_i$, and $v(\by,\t)\equiv V(\bx,t)$. We can
rewrite the PDDE as follows.
\[
-\pda{v(\by,\t)}{\t} - \frac{\m_0\a_1-\m_1\a_0}{\a_1-\a_0}v(\by,\t) -
\frac{\m_1-\m_0}{\a_1-\a_0}v(\bal\by,\t) = 0
\]
To get rid of the second term we multiply the PDDE with an integrating
factor. This then leads to 
\begin{equation}
\pda{f(\by,\t)}{\t} = \tl f(\bal\by,\t) = f(\bbe\by,\t)
\label{eq:12}
\end{equation}
where $f(\by,\t)\equiv
v(\by,\t)\,\smash{\exp(\frac{\m_0\a_1-\m_1\a_0}{\a_1-\a_0}\t)}$,
$\tl\equiv-\smash{\frac{\m_1-\m_0}{\a_1-\a_0}}$, and
$\bbe\equiv\tl\bal$. We used homogeneity to pull $\tl$ inside the
function $f$. As it will turn out, the $\bbe$ are the crucial parameters
in the pricing problem. Using the ansatz
\begin{equation}
\label{eq:13}
f(\by,\t) = \sum_{n\ge0}w_n(\t)g\big((\b_0)^ny_0,(\b_1)^ny_1\big) =
\sum_{n\ge0}g\big(w_n(\b_0\t)y_0,w_n(\b_1\t)y_1\big)
\end{equation}
it is easy to prove that Eq.~(\ref{eq:13}), with
$w_n(\t)=\frac{\t^n}{n!}$ and $g(\by)=v(\by,0)$, solves the PDDE in
Eq.~(\ref{eq:12}). Putting everything back together, the solution
becomes
\begin{equation}
V(\bx,t) = \sum_{n\ge0} V\bigg( x_0
\frac{(\b_0\t)^n}{n!}e^{-\b_0\t}, x_1
\frac{(\b_1\t)^n}{n!}e^{-\b_1\t},T\bigg)
\end{equation}
Note that the parameters describing the dynamics only enter the
solution via $\bbe$. These objects are numeraire invariant as can be
seen easily by inspection, using $\m_i\to\m_i-\k,\a_i\to\a_i/\g$. Now
it also becomes completely obvious that in order to avoid arbitrage,
we have to restrict the $\bbe$ to be positive, or equivalently
$\tl>0$ since $\a_i>0$. Otherwise an option with positive payoff at
maturity could have a negative value at an earlier time. In the
present example, this leads to the following restriction, as we
mentioned before
\[
\tl = -\frac{\m_1-\m_0}{\a_1-\a_0}>0
\]
Finally note that there is no reference to the intensity of the
real world process $N_t$. It completely drops out of the problem. This
should indeed be the case since we have hedged our position and once a
perfect hedge has been found which annihilates jump risk, we become
indifferent to the frequency with which jumps occur. Note that
this result still holds if the jump intensity in the real world
measure would be a stochastic process. This does not mean that the
real world intensity does not play any role at all in pricing. In
practice, the jump intensity is closely related to the magnitude of
the drift terms, which do enter the price formulas.

\subsubsection{An interpretation}
\label{sec:an-interpretation}

Let us try to connect this result to the usual formulation in terms
of martingales. By using homogeneity, we can rewrite Eq.~(\ref{eq:13}) as
\[
V(\bx,t) = \sum_{n\ge0} \frac{(\b_0\t)^n}{n!}e^{-\b_0\t}
V\bigg( x_0, x_1 \bigg(\frac{\b_1}{\b_0}\bigg)^n
e^{-(\b_1-\b_0)\t}, T \bigg)
\]
In this form, the formula can be interpreted as an expectation
under a Poisson process with intensity $\b_0$, where $x_1$
makes jumps of size $\b_1/\b_0=\a_1/\a_0$ and has a drift
$-(\b_1-\b_0)$ relative to $x_0$. Now let's assume that $x_0$
is a `risk-free' bond\footnote{Of course, the notion of a 'risk-free'
asset is defined relative to some numeraire, usually money, and
means that under this numeraire the dynamics of that particular
asset is locally deterministic.}.
This corresponds to setting $\a_0=1$ and $\m_0=r$. Further assume
that $x_1$ is a risky bond, making downward jumps, so $\a_1=\a<1$ and
$\m_1=\m$. Then
\[
  \b_0 = \tl \quad,\quad 
  \b_1 = \tl\a \quad,\quad
  \tl = -\frac{\m-r}{\a-1}
\]
The no-arbitrage condition now amounts to $\m>r$. This has a nice
intuitive interpretation: any downward jumprisk should be compensated
for by a higher return than the riskfree rate (this is known as the
credit spread). In fact we can rewrite the definition of $\tl$ as follows
\[
\m=r+\tl(1-\a)\ge r
\]
This suggest that $\tl(1-\a)=\b_0-\b_1$ denotes the credit spread due
to default risk and $\tl$ could be interpreted as the 'market price of
default risk', as was already noted in
Ref.~\cite{VellekoopBeumeeHilberink99}. One should however be careful
to attach an invariant meaning to the latter, since it is not
numeraire invariant. Indeed, if we change the numeraire to $x_1$,
which corresponds to setting $\k=\m$ and $\g=\a$ in Eq.~(\ref{eq:11}),
we find that $\tl\rightarrow\a\tl$. Also, in this numeraire it would
be more natural to rewrite Eq.~(\ref{eq:13}) as follows
\[
V(\bx,t) = \sum_{n\ge0} \frac{(\b_1\t)^n}{n!}e^{-\b_1\t}
V\bigg( x_0 \bigg(\frac{\b_0}{\b_1}\bigg)^n
e^{-(\b_0-\b_1)\t}, x_1, T \bigg)
\]
But in this form, the formula can be interpreted as an expectation
under a Poisson process with intensity $\b_1$, where $x_0$
makes jumps of size $\b_0/\b_1=\a_0/\a_1=1/\a>1$ and has a drift
$-(\b_0-\b_1)$ relative to $x_0$. So we recover the well known fact
that the choice of a martingale measure is also relative to the choice
of a numeraire. Only the symmetric formula Eq.~(\ref{eq:13}) is
canonical and therefore, in our view, more fundamental.

\subsection{The general case}
\label{sec:general-case}

Let us now consider the general case, where the price processes of
the tradables are driven by both Wiener and Poisson processes.
To this end we consider $N+1$ tradables with price processes
$X^i_t\,(i=0,\ldots,N)$ of the form
\begin{equation}
\label{eq:14}
  \frac{dX^i_t}{X^i_{t_-}} 
  = \m_i(\bX_{t_-},t)\, dt 
  + \msum \s^m_i(\bX_{t_-},t)\, dW^m_t
  + \rsum \big(\a^r_i(\bX_{t_-},t)-1\big)\, dN^r_t
\end{equation}
where $\{ (W_t,\cF_t),\, t\geq 0\}$ and $\{ (N_t,\cF_t),\, t\geq 0\}$
are $M$ standard Wiener and $R$ Poisson processes with $M+R=N$, all
independent of each other, and we assume that the Poisson processes
all have a strictly positive intensity for all $t\geq 0$. We will
again find that the actual values of the intensities do not play any
role in the pricing equations. The functions $\m_i(\bx,t)$,
$\s^m_i(\bx,i)$ and $\a^r_i(\bx,t)$ are deterministic, known in
advance and should be homogeneous of degree zero in the tradables.
They will be restricted by no-arbitrage constraints. In order to
keep the notation transparent, we will omit the parameters of these
functions in what follows.

\subsubsection{The first step: invoking It\^o's rule}
\label{sec:change-variables}

We want to compute the price of a derivative security whose price
$V(\bX_t,t)$ depends on tradables satisfying Eq.~(\ref{eq:14}). For
any function $V(\bx,t)$ for which the second order derivatives in
$\bx$ and the first order derivative in $t$ exist, we can write the
generalized It\^o rule:
\begin{eqnarray*}
\lefteqn{V(\bX_t,t) - V(\bX_0,0) = }&&\\
&=&
\int_0^t \pda{V(\bX_{s-},s)}{s}\, ds
+ \isum\int_0^t \pda{V(\bX_{s-},s)}{x_i}\, dX^i_s
\\
&& 
+ \shalf \sum_{i,j=0}^N
\int_0^t \pdb{V(\bX_{s-},s)}{x_i}{x_j}\, 
d\big[X^i, X^j\big]^c_s 
\\
&&
+ \sum_{0<s\leq t} \left( V(\bX_s,s)-V(\bX_{s-},s) 
- \sum_{i=0}^N \pda{V(\bX_{s-},s)}{x_i}\, \Delta X_s^i \right)
\end{eqnarray*}
where $\{ [X^i,X^j]^c_t,\, t\geq 0\}$ is the continuous part of the
quadratic covariation process associated with the processes $X^i_t$
and $X^j_t$ and $\D X_s^i\equiv X^i_s-X^i_{s_-}$.  Note that all functions only
need to be evaluated on the left-continuous part of the process. In
short-hand notation and by obvious substitutions this means that
\begin{eqnarray*}
\lefteqn{dV(\bX_t,t) =}
\\
&=&
\sum_{i=0}^N X^i_{t_-}\pda{V(\bX_{t_-},t)}{x_i}\,
\bigg(
  \m_i\, dt + \sum_{m=1}^M \s_i^m dW_t^m+ \sum_{r=1}^R
  \big(\a_i^r-1\big)\, dN_t^r 
\bigg)
\\
&&
+ \bigg(
\pda{V(\bX_{t_-},t)}{t}
+ \shalf\sum_{i,j=0}^N 
\sum_{m=1}^M\s_i^m\s_j^m\, X^i_{t_-} X^j_{t_-} \pdb{V(\bX_{t_-},t)}{x_i}{x_j}
\bigg)\, dt
\\
&&
+ \sum_{r=1}^R \bigg( 
V\big(\bal^r \bX_{t_-},t\big) - V(\bX_{t_-},t) 
-\sum_{i=0}^N \big(\a_i^r-1\big) X^i_{t_-}\pda{V(\bX_{t_-},t)}{x_i}
\bigg)\, dN_t^r
\end{eqnarray*}
where $\bal^r \bX_{t_-}\equiv \big(\a^r_0X^0_{t_-},\ldots,\a^r_N
X^N_{t_-}\big)$. Now, just as in the previous cases, we want to 
use this equation to derive conditions under which a given portfolio
is self-financing. To this end, we add terms of the form
\[
\phi_i(\bX_{t_-},t)\bigg(dX^i_t-\m_i X^i_{t_-}\, dt 
+ \sum_{m=1}^M \s^m_i X^i_{t_-}\, dW^m_t
+ \sum_{r=1}^R \big(\a^r_i-1\big) X^i_{t_-}\, dN^r_t\bigg)
\]
which vanish by virtue of the dynamic equations. The functions 
$\phi_i(\bX_{t_-},t)$ constitute hedge ratios which are yet to be
determined. They must only depend on the left-continuous processes
$\bX_{t_-}$ because we cannot allow our hedging strategy to anticipate
the sudden jumps. The dynamic equation for $V$ now becomes
\begin{eqnarray}
\lefteqn{dV(\bX_t,t) =}\nn\\
&=&
\sum_{i=0}^N 
\phi_i(\bX_{t_-},t)\, dX_t^i 
\nn
\\
&& 
+ \bigg\{
\pda{V(\bX_{t_-},t)}{t} 
+ \shalf\sum_{i,j=0}^N \sum_{m=1}^M
\s_i^m\s_j^m\, X^i_{t_-}X^j_{t_-}\pdb{V(\bX_{t_-},t)}{x_i}{x_j}
\nn
\\
&&
\quad\quad
+ \sum_{i=0}^N \m_i X^i_{t_-}
\bigg(\pda{V(\bX_{t_-},t)}{x_i}-\phi_i(\bX_{t_-},t)\bigg)
\bigg\}\, dt 
\nn
\\
&& 
+ \sum_{m=1}^M \bigg\{
\sum_{i=0}^N \s_i^m X^i_{t_-} 
\bigg( \pda{V(\bX_{t_-},t)}{x_i}-\phi_i(\bX_{t_-},t)\bigg) 
\bigg\}\, dW_t^m
\nn
\\
&&
+ \sum_{r=1}^R \bigg\{
V\big(\a^r \bX_{t_-},t\big) 
- V(\bX_{t_-},t)
\nn
\\
&&
\quad\quad%\hspace{2cm}
- \sum_{i=0}^N \big(\a_i^r-1\big) X^i_{t_-}\phi_i(\bX_{t_-},t)
\bigg\}\, dN_t^r
\nn
\\
&\stackrel{\rm def}{=}& 
\sum_{i=0}^N \phi_i(\bX_{t_-},t)\, dX_t^i 
+ \cL V(\bX_{t_-},t)\, dt 
\nn
\\
&&
+ \msum\cL_W^m V(\bX_{t_-},t)\,dW_t^m 
+ \rsum\cL_N^r V(\bX_{t_-},t)\,dN_t^r
\label{eq:15}
\end{eqnarray}
where we defined
\begin{eqnarray*}
\cL V(\bx,t) 
&\equiv&
\pda{V(\bx,t)}{t}
+ \shalf\sum_{i,j=0}^N \sum_{m=1}^M
\s_i^m\s_j^m\, x_ix_j\pdb{V(\bx,t)}{x_i}{x_j}
\\
&&+ \sum_{i=0}^N \m_i x_i
\bigg(\pda{V(\bx,t)}{x_i}-\phi_i(\bx,t)\bigg),
\end{eqnarray*}
\[
  \cL_W^m V(\bx,t) \equiv\isum \s_i^m x_i 
  \bigg( \pda{V(\bx,t)}{x_i}-\phi_i(\bx,t)\bigg) 
\]
for $m=1,\ldots,M$ and
\[
  \cL_N^r V(\bx,t) \equiv V\big(\bal^r \bx,t\big) - V(\bx,t)
  - \isum\big(\a_i^r-1\big) x_i\phi_i(\bx,t)
\]
for $r=1,\ldots,R$.

\subsubsection{The self-financing portfolio conditions}
\label{sec:self-financ-portf}

From Eq.~(\ref{eq:15}) it is now obvious that a portfolio consisting of
$\phi_i(x,t)$ assets $i$ at time $t$ will be self-financing if and
only if
\begin{eqnarray}
  \label{eq:16}
  \cL V(\bx,t) &=& 0
  \\
  \label{eq:17} 
  \cL_W^m V(\bx,t) &=& 0 \qquad\forall\, m=1,\ldots,M
  \\
  \label{eq:18}
  \cL_N^r V(\bx,t) &=& 0 \qquad\forall\, r=1,\ldots,R
\end{eqnarray}
while at the same time,
\[
V(\bx,t) = \sum_{i=0}^N \phi_i(\bx,t)\, x_i
\]
Using homogeneity this last condition can be shown to be equivalent to
\begin{equation}
  \label{eq:19}
  \sum_{i=0}^N x_{i} \bigg( \pda{V(\bx,t)}{x_i}-\phi_{i}(\bx,t) \bigg) = 0
\end{equation}
Note that it is easy to see that in the absence of jump processes,
we must have $\phi_i(\bx,t)=\smash{\pda{V(\bx,t)}{x_i}}$, i.e. we
retrieve the usual Black-Scholes delta hedge. Now we use
Eq.~(\ref{eq:19}) and homogeneity to rewrite Eq.~(\ref{eq:18})
in a slightly different form, which will turn out to be useful
when solving the constraints
\begin{eqnarray}
  0 
  &=&
  \chi^r(\bx,t) 
  \nn
  \\
  &=& 
  V\big(\bal^r \bx,t\big)-V(\bx,t)
  \nn
  \\
  &&
  + \sum_{i=0}^N (\a_i^r-1) x_i\bigg(\pda{V(\bx,t)}{x_i}-\phi_i(\bx,t)\bigg)
  \nn
  \\
  &&
  + \sum_{i=0}^N x_i\pda{V(\bx,t)}{x_i} 
  - \sum_{i=0}^N \a_i^r x_i\pda{V(\bx,t)}{x_i}
  \nn
  \\
  &=&
  \sum_{i=0}^N \a_i^rx_i\bigg(\pda{V(\bx,t)}{x_i}-\phi_i(\bx,t)\bigg) 
  - \sum_{i=0}^N \a_i^r x_i\pda{V(\bx,t)}{x_i}
  + V(\bal^r \bx,t) 
  \label{eq:20}
\end{eqnarray}
Indeed, collecting Eqs.~(\ref{eq:17}),(\ref{eq:19}), and
(\ref{eq:20}), we now have a set of $M+R+1=N+1$ linear equations for
the $N+1$ unknown hedge ratios $\phi_i(x,t)$:
\begin{equation}
  \label{eq:21}
  \begin{pmatrix}
    1      & \cdots & 1\\
    \s_0^1 & \cdots & \s_N^1 \\
    \vdots & \ddots & \vdots \\
    \s_0^M & \cdots & \s_N^M \\
    \a_0^1 & \cdots & \a_N^1 \\
    \vdots & \ddots & \vdots \\
    \a_0^R & \cdots & \a_N^R
  \end{pmatrix}
  \begin{pmatrix}
    x_0 \big( \spda{V}{x_0}-\phi_0\big) \\
    \vdots\\
    \vdots\\
    \vdots \\
    x_{N} \big( \spda{V}{x_N}-\phi_N \big) \\
  \end{pmatrix}
  =
  \begin{pmatrix}
    0\\
    \vdots\\
    0\\
    \sum_{i=0}^N \a_i^1x_i\spda{V}{x_i}
    -V\big(\bal^1\bx,t\big)\\
    \vdots\\
    \sum_{i=0}^N \a_i^R x_i\spda{V}{x_i}
    - V\big(\bal^R \bx,t\big)\\
  \end{pmatrix}
\end{equation}
Under obvious non-singularity conditions this has a unique solution.
We will denote the matrix in the above equation by $A$. Substituting
the solution in the remaining Eq.~(\ref{eq:16}), we find
\begin{eqnarray}
\cL V(\bx,t)
  &=&
  \pda{V(\bx,t)}{t} + \shalf\sum_{i,j=0}^N
  \sum_{m=1}^M\s_i^m\s_j^m x_i x_j \pdb{V(\bx,t)}{x_i}{x_j}
  \nn
  \\
  &&
  + \sum_{i=0}^N \m_i x_i\bigg(\pda{V(\bx,t)}{x_i}-\phi_i(\bx,t)\bigg)
  \nn
  \\
  &=& 
  \pda{V(\bx,t)}{t} + \shalf
  \sum_{i,j=0}^N\sum_{m=1}^M \s_i^m\s_j^mx_ix_j\pdb{V(\bx,t)}{x_i}{x_j}
  \nn
  \\
  &&
  + \sum_{r=1}^R \bigg(
  V\big(\bbe^r \bx,t\big) 
  -\sum_{i=0}^N\b^r_i x_i\pda{V(\bx,t)}{x_i}\bigg).  
  \label{eq:22}
\end{eqnarray}
where we define quantities
\begin{equation}
  \label{eq:23}
  \b^r_i \equiv \a^r_i\tl_r, \hspace{5mm}
  \tl_r \equiv -\sum_{j=0}^N \m_j A_{j,M+r}^{-1} 
\end{equation}
Again, we used homogeneity to pull the terms $\tl_r$ inside
the function $V$ using $V(\tl_r\bal^r\bx,t)=V(\bbe^r\bx,t)$. 
At this point we have found the PDDE which the price of a
derivative security must satisfy. It now remains to show that
this PDDE respects the numeraire invariance. In fact, we will
proceed to show that the $\b^r_i$ are themselves numeraire
invariant quantities, and (in Section~\ref{sec:power-tradables})
that the no-arbitrage condition requires them to be positive.

\subsubsection{Symmetries of the PDDE}
\label{sec:symmetries-pdde}

By now, it should not surprise the reader that the pricing
PDDE Eq.~(\ref{eq:22}) respects numeraire invariance, since
we explicitly used this fact in its derivation. Nevertheless
we will once more prove this property, as a consistency check.
In the present model, a general change of numeraire consists of:
\begin{itemize}
\item a shift of the drift-terms: $\m_i\to\m_i-\k$
\item shifts of the volatility functions: $\s^m_i\to\s^m_i-g_m$
\item the scaling of the jump sizes: $\a^r_i\to\a^r_i\g_r$
\end{itemize}
The proof of the invariance of the diffusion part of the
PDDE is completely identical to the derivation in
Section~\ref{sec:symmetries-pde}. The invariance of the jump
part of the equation requires a little extra work. Let us
consider Eq.~(\ref{eq:23}) again
\[
  \b^r_i \equiv \a^r_i\tl_r, \hspace{5mm}
  \tl_r \equiv -\sum_{j=0}^N \m_j A_{j,M+r}^{-1} 
\]
The $\tl_r$ can be viewed as part of a vector ${\bf v}
\equiv(v_0,\varphi_1,\ldots,\varphi_M,-\tl_1,\ldots,-\tl_R)$, which
is a solution of
\begin{equation}
  \label{eq:24}
  \m_j
  = \sum_{i=0}^N v_i A_{ij}
  = v_0
  + \sum_{m=1}^M \varphi_m \s_j^m
  - \sum_{r=1}^R \tl_r \a_j^r
\end{equation}
Obviously, if we can prove the invariance of the $\bbe^r$,
invariance of the PDDE follows. In fact we will show
a little more, namely that the $\varphi_m$ are also numeraire
invariant. First we consider the effect of scaling the
jumpsizes $\a^r_i\to\a^r_i\g_r$. Then the matrix components
$A_{ij}$ transform like
\[
  A_{ij} \to D_i A_{ij}
\]
where $D\equiv\big(1,\ldots,1,\g_1,\ldots,\g_r\big)$.
This automatically implies that we should have $v_i\to D^{-1}_i v_i$
and thus $\tl_r\to\tl_r\g_r^{-1}$ and $\varphi_m\to\varphi_m$.
Since $\b^r_i\equiv\a^r_i\tl_r$, we therefore have
\[
  \b^r_i\to (\a^r_i\g_r)(\tl_r\g_r^{-1})=\a^r_i\tl_r=\b^r_i
\]
Thus the $\bbe^r$ and $\varphi_m$ are invariant under scaling of
the jump sizes. Now consider what happens when all $\m_j$ are
shifted by $-\k$. In this case we need a correction on ${\bf v}$
such that
\[
  \m_j-\k = \sum_{i=0}^N (v_i+\d v_i)A_{ij}
\]
is satisfied. The particular form of $A$ shows that this can only be
done by setting $\d{\bf v}=(-\k,0,\ldots,0)$. But this shows that the
$v_i$ with $i\neq 0$ are invariant under the shifts, and the same
holds for $\bbe^r$ and $\varphi_m$. Note that we effectively use
the portfolio replication condition Eq.~(\ref{eq:19}). In a similar
manner we can show the invariance under shifts $\s^m_i\to\s^m_i-g_m$
in the volatility functions. Indeed, we now need a correction such that
\[
  \m_j = \sum_{i=0}^N (v_i+\d v_i) A_{ij} 
       - \sum_{m=1}^M (v_m+\d v_m) g_m
\]
is satisfied. Since the terms depending on $g_m$ do not depend on the
index $j$, it is not hard to see that the proper choice is
\[
  \d{\bf v} = \bigg( \sum_{m=1}^M v_m g_m,0,\ldots,0\bigg)
\]
again showing the invariance of the $\bbe^r$ and $\varphi_m$ in this case.

\subsubsection{Market prices of risk}
\label{sec:market-prices-risk}

The `market prices of risk' is a notion that is often used to
indicate the willingness of the market to hold
a risky asset compared to holding a risk-free asset. In the case of
jump-diffusion more care should be taken when introducing such
concepts.  In particular we show that the usual definition of
`market-price of risk' for a Poisson process is not a numeraire
invariant quantity. This is in contrast to the case of Wiener
processes. Thus when talking about `market price of risk' for a
Poisson process, one should always make clear what particular
numeraire is chosen. However, it is possible to formulate everything in
terms of numeraire invariant quantities. This is exactly why we
introduced the $\bbe^r$. Expressed in these objects no confusion can
arise to what is meant. Let us once more consider Eq.~(\ref{eq:24})
\[
\m_i = v_0+\sum_{m=1}^M \varphi_m\s^m_i-\sum_{r=1}^R \tl_r\a^r_i
\]
As we have seen in the previous section, both left- and right-hand side
are not numeraire invariant quantities. But it is straightforward
to introduce a quantity which is numeraire invariant and tells us
something about the `market prices of risk'. Indeed, consider the
difference $\m_i-\m_j$
\begin{equation}
\label{eq:25}
  \m_i-\m_j=\sum_{m=1}^M \varphi_m\big(\s^m_i-\s^m_j\big)
           -\sum_{r=1}^R \tl_r\big(\a^r_i-\a^r_j\big)
\end{equation}
This is the general, numeraire invariant, expression that provides a
relation between the returns of the tradables, their volatilities, and
jump sizes. The quantities that, in the literature, are called `market
prices of risk' are $\varphi_m$ and $\tl_r$. Note however that only
the $\varphi_m$ of these are numeraire invariant. The $\tl_r$ are
definitely not. This is a strong argument in favor of the
$\b^r_i=\a^r_i\tl_r$ as the fundamental quantities for Poisson
processes, since they are, like the $\varphi_m$, numeraire invariant.
It is for this reason, that in the remainder of this article we will
only use the $\b^r_i$.

\paragraph{The usual formulation}
\label{sec:usual-formulation}

The usual formulation of `market price of risk' amounts to the choice
of money as a numeraire. Next we pick a tradable, say tradable $0$,
that we call `risk-free' with drift $r$, no volatility $\s^m_0=0$, and
no jumps, $\a^r_0=1$.  Then Eq.~(\ref{eq:25}) reduces to (with $j=0$
and $i=1,\ldots,N$)
\begin{equation}
\label{eq:26}
  \m_i = r + \sum_{m=1}^M \varphi_m\s^m_i
           - \sum_{r=1}^R \tl_r\big(\a^r_i-1\big)
\end{equation}
For the two simplest cases we can then easily write down the
corresponding `market prices of risk'. For the case where $M=1,R=0$
this amounts to (dropping indices)
\[
\varphi = \frac{\m-r}{\s}
\]
and in the case where $M=0,R=1$ we get
\[
\tl = \frac{\m-r}{1-\a}
\]

\subsubsection{A general solution}
\label{sec:general-solution}

In this section we present a solution of the PDDE $\cL V=0$,
with $\cL V$ as given in Eq.~(\ref{eq:22}), for the case where
the $\m_i(t)$ and $\s^m_i(t)$ only depend on $t$ and the $\a^r_i$
are constant. We consider a European style derivative security, where
a payoff $V(\bx,T)$ is specified as boundary condition at maturity
$T$. In appendix~\ref{sec:time-dependent-case} we proof that the
price of such a contract at time $t<T$ is given by
\begin{eqnarray}
  \lefteqn{V(\bx,t) =}
  \nn
  \\
  &&
  \sum_{\bn\ge0}
  \int_{\bR^M}
  V\bigg(
  x_0\phi(\bz-\bth_0)\p_\bn(\brh_0),\ldots,
  x_N\phi(\bz-\bth_N)\p_\bn(\brh_N),
  T\bigg)d\bz
  \label{eq:27}
\end{eqnarray}
where $\bz\equiv(z_1,\ldots,z_M)$, $\bn\equiv(n_1,\ldots,n_R)$,
and $\bn\ge0\equiv\{\bn|n_1\ge0,\ldots,n_R\ge0\}$. 
The $N+1$ vectors $\bth_i\equiv(\theta^1_i,\ldots,\theta^M_i)$
have dimension $M$. They are found from a singular value
decomposition of the time-integrated covariance matrix (which
has full rank, since $A$ is assumed to be invertible)
\[
  \sum_{m=1}^M \theta_i^m(t)\theta_j^m(t)
  \equiv \int_t^T \sum_{m=1}^M \s_i^m(u)\,\s_j^m(u)\,du
\]
The $N+1$
vectors $\brh_i\equiv(\r^1_i,\ldots,\r^R_i)$
are defined by
\[
  \r_i^r(t)\equiv\int_t^T\b^r_i(u)\,du
\]
The functions $\phi(\bz)$ and $\p_\bn(\brh_i)$ are defined by
\begin{eqnarray*}
  \phi(\bz) &\equiv&
  \prod_{m=1}^M\frac{1}{\sqrt{2\p}}\,e^{-\shalf z_m^2}
\\
  \p_\bn(\brh_i) &\equiv&
  \prod_{r=1}^R\bigg(\frac{(\r_i^r)^{n_r}}{n_r!}\,
  e^{-\r_i^r}\bigg)
\end{eqnarray*}
Note that the general solution Eq.~(\ref{eq:27}) is in a symmetric,
canonical form. It is form invariant under numeraire changes.
Of course, one could relate this solution to an expectation under
some equivalent martingale measure, by choosing a numeraire and
using homogeneity to bring some functions out of $V$, as we have shown
in Section~\ref{sec:an-interpretation}. However, we prefer the
symmetric form, since it makes the numeraire invariance manifest.

\subsubsection{The single pure jump process revisited}
\label{sec:single-pure-jump}

Let us reconsider the example of Section~\ref{sec:simple-example},
which corresponds to the case $R=1$, $M=0$. The price processes
for the two tradables are given by
\[
  \frac{dX^i_t}{X^i_{t_-}} = \m_i\, dt + (\a_i-1)\,dN_t,
  \hspace{5mm} (i=0,1)
\]
where the $\a_i$ and $\m_i$ are constant and $\a_i>0$.
The corresponding matrix $A$ of constraints is given by
\[
A =
\begin{pmatrix}
  1 & 1 \\
  \a_0 & \a_1
\end{pmatrix}
\]
For $A$ to be invertible we need to have $\a_0\ne\a_1$.
The corresponding $\bbe$ are
\[
  \b_0 = -\a_0\frac{\m_1-\m_0}{\a_1-\a_0} \quad,\quad \b_1 =
  -\a_1\frac{\m_1-\m_0}{\a_1-\a_0}
\]
as expected. The no-arbitrage conditions $\b_i>0$ become
$\m_0<\m_1$ iff $\a_0>\a_1$. Since the $\bbe$ are
constant, we readily write down the price of a European contract
at time $t$, given the payoff $V(\bx,T)$ at maturity $T$,
using Eq.~(\ref{eq:27})
\[
  V(\bx,t) = \sum_{n\ge0} V\big(
  x_0\p_n(\b_0\t),x_1\p_n(\b_1\t),T\big)
\]
It is now a simple matter to determine the equations for
the hedge ratios $\phi_i(\bx,t)$ in terms of the two tradables,
\[
\begin{pmatrix}
  x_0 \big( \spda{V}{x_0}-\phi_0 \big) \\
  x_1 \big( \spda{V}{x_1}-\phi_1 \big)
\end{pmatrix}
=
\begin{pmatrix}
  1 & 1 \\
  \a_0 & \a_1
\end{pmatrix}^{-1}
\,
\begin{pmatrix}
  0\\
  \sum_{i=0,1}\a_i x_i \spda{V}{x_i} - V(\bal \bx,t)
\end{pmatrix}
\]
Solving for $\phi_i(\bx,t)$ we get
\begin{eqnarray*}
  \phi_0(\bx,t)
  &=& 
  \frac{\a_1\sum_{i=0,1} x_i\spda{V(\bx,t)}{x_i}
  -V(\bal\bx,t)}{x_0(\a_1-\a_0)} 
  \quad=\quad
  \frac{\a_1 V(\bx,t)-V(\bal\bx,t)}{x_0(\a_1-\a_0)}
  \\
  \phi_1(\bx,t)
  &=& 
  \frac{\a_0 \sum_{i=0,1} x_i\spda{V(\bx,t)}{x_i}
  -V(\bal\bx,t)}{x_1(\a_0-\a_1)} 
  \quad=\quad
  \frac{\a_0 V(\bx,t)-V(\bal\bx,t)}{x_1(\a_0-\a_1)}
\end{eqnarray*}
where we have used homogeneity again.

\subsubsection{Power tradables}
\label{sec:power-tradables}

In our approach we stress the importance of using proper coordinates
to formulate pricing problems. We showed that this leads to
the use of tradables, or self-financing objects as coordinates.
Depending on the type of pricing problem it may be useful to introduce
additional tradables, derived from the initial ones, to simplify
calculations. In Ref.~\cite{HooglandNeumann99b} we introduced
so-called power-tradables to derive compact expressions for exotic
options in a diffusion setting. We can do similar things in the
case of jump-diffusion processes. To this end consider tradables with
the following payoff at maturity $T$:
\[
V\big(x_0,\ldots,x_N,T\big) = \prod_{i=0}^N x_i^{\eta_i}
\]
where we have to impose the constraint $\isum\eta_i=1$ in order
to get a homogeneous function of degree one. Now it is a simple matter to
compute the value of a tradable with this payoff at an earlier time
in the context of Sec.~\ref{sec:general-solution}. Working out
Eq.~(\ref{eq:27}) for this case we arrive at
\[
V\big(x_0,\ldots,x_N,t\big) = e^{\xi(t)} \prod_{i=0}^N x_i^{\eta_i}
\]
where
\[
 \xi(t) = -\frac{1}{2}\sum_{m=1}^M\sum_{i<j} 
            \eta_i\eta_j(\theta_i^m-\theta_j^m)^2
          +\sum_{r=1}^R\left(\prod_{i=0}^N(\r_i^r)^{\eta_i}
          -\sum_{i=0}^N \eta_i\r_i^r\right)
\]
By using the relations $\smash{x_i\pda{V(\bx,t)}{x_i}=\eta_iV(\bx,t)}$
and $\smash{V(\bal^r\bx,t)=\prod_{i=0}^N (\a_i^r)^{\eta_i}V(\bx,t)}$
we find from Eq.~(\ref{eq:21}) that the hedge ratios should satisfy
\[
  A\begin{pmatrix} x_0\phi_0 \\ \vdots \\ x_N\phi_N\end{pmatrix} =
  \begin{pmatrix} 1 \\ \psi^1 \\ \vdots \\ \psi^M \\ \chi^1
  \\ \vdots \\ \chi^R \end{pmatrix} V(\bx,t), \hspace{5mm}
  \psi^m = \isum \s_i^m \eta_i, \hspace{5mm}
  \chi^r = \prod_{i=0}^N (\a_i^r)^{\eta_i}
\]
From this we derive the dynamic equations for the power tradable
\begin{eqnarray*}
  \frac{dV_t}{V_{t_-}} &=& \isum \phi_i \frac{dX^i_t}{V_{t_-}}\\
  &=& \left(v_0+\msum\psi^m\varphi_m-\rsum\chi^r\tl_r\right)dt
  +\msum\psi^m dW^m_t+\rsum (\chi^r-1)dN^r_t
\end{eqnarray*}
where $v_0,\varphi_m,\tl_r$ were defined in Eq.~(\ref{eq:24}).
One important use of the power tradables is, that they allow
us to construct a new basis in the space of tradables in which
the matrix $A$ is `diagonal' in the sense that every tradable
depends on at most one stochastic factor. This amounts to finding
a proper set of $\eta_i$ given a set of values for $\psi^m,\chi^r$.
But this is just a linear set of equations. Indeed, taking the
log of $\chi^r$ we see that the $\eta_i$ should satisfy
\[
  \isum\eta_i=1,\hspace{5mm}
  \isum\s_i^m\eta_i=\psi^m, \hspace{5mm}
  \isum\log(\a_i^r)\eta_i=\log(\chi^r)
\]
This means that under obvious non-singularity conditions, it
is possible to construct a new set of tradables ${\bf Y}$ satisfying
\[
  \frac{dY^0_t}{Y^0_{t_-}} = 0, \hspace{5mm}
  \frac{dY^m_{W,t}}{Y^m_{W,t_-}} = \varphi^m dt + dW^m_t, \hspace{5mm}
  \frac{dY^r_{N,t}}{Y^r_{N,t_-}} = -\tl^r dt + dN^r_t
\]
under the numeraire $Y^0$. This gives a very nice illustration of the
concept of market price of risk under this numeraire. On the other
hand, it clearly shows that the $\tl^r$, and consequently the $\b_i^r$,
should be positive (almost surely) in order to avoid arbitrage in
the model. Indeed, otherwise one of the $Y^r$ would be strictly
increasing relative to $Y^0$, indicating an arbitrage opportunity.
These no-arbitrage conditions still hold in the more general setting,
but we will not explicitly derive them here. Finally note that power
tradables are closely related to optimal growth or Kelly strategies.
For more applications of power tradables we refer to
Ref.~\cite{HooglandNeumann99b}.

\subsubsection{Levy processes: the rough guide}

In this section we sketch how our model could be used in a simple
example of an incomplete market, and, in a limit, in a market
in which the price is driven by a Levy process. We start by defining
a market consisting of two tradables, a stock $S$ and a bond $P$.
Taking the bond as numeraire, we assume that the stockprice is driven
by one Wiener and $R$ Poisson processes. So the dynamic equations
take the following form
\[
  \frac{dS_t}{S_{t_-}} = \m\, dt + \s dW_t
  +\sum_{r=1}^R \big(\a^r-1\big)\, dN^r_t
\]
\[
  \frac{dP_t}{P_{t_-}}=0
\]
Obviously, this market is not complete. However, we can make
it complete by introducing $R$ extra tradables $P^r$, as follows
\[
  \frac{dP^r_t}{P^r_{t_-}} = \m_r\, dt
  +\big(\a^r-1)\, dN^r_t
\]
The essential freedom that we have is the choice of the drift
terms $\m_r$. We will only assume that they are constant, and
such that the total market is arbitrage free. Now we can apply
the general theory from the previous sections. Under the
identification $S=X^0, P=X^1, P^r=X^{r+1}$ the matrix $A$
becomes
\[
  A = \left( \begin{array}{cccccc}
  1 &  1 & 1 & 1 & \cdots & 1 \\
  \s & 0 & 0 & 0 & \cdots & 0 \\
  \a^1 & 1 & \a^1 & 1 & \cdots & 1 \\
  \a^2 & 1 & 1 & \a^2 & \cdots & 1 \\
  \vdots & \vdots & \vdots & \vdots & \ddots & \vdots \\
  \a^R & 1 & 1 & 1 & \cdots & \a^R
  \end{array} \right)
\]
For the $\b^r_i$ we find the following expressions
\[
  \tl_r = \frac{\m_r}{1-\a^r}, \hspace{5mm}
  \b_i^r = \left\{ \begin{array}{ll}
  \a^r\tl_r & \mbox{if $i=0$ or $i=r+1$}\\
  \tl_r     & \mbox{otherwise}
  \end{array} \right.
\]
From this we see that the no-arbitrage conditions are simply
$\m_r>0$ iff $\a^r<1$ for all $r$. Now using the identity
$\b^r_i=\b^1_i-\m_r(\d_{i,0}+\d_{i,r+1})$ and homogeneity, the
pricing PDDE takes the following form
\begin{eqnarray*}
  \pda{V(\bx,t)}{t} &+& \shalf\s^2 S^2
    \frac{\partial^2V(\bx,t)}{\partial S^2}\\
      &+& \sum_{r=1}^R \bigg(
  V(\bbe^r\bx,t)-V(\bbe^1\bx,t)
  +\m_rS\frac{\partial V(\bx,t)}{\partial S}
  +\m_rP^r\frac{\partial V(\bx,t)}{\partial P^r}\bigg)
\end{eqnarray*}
Since our original market only contains $S$ and $P$, we will
only be interested in contracts which are specified in terms
of only these two tradables. It is not hard to see that in this
case the PDDE can be reduced to
\begin{eqnarray*}
  \pda{V(S,P,t)}{t} &+& \shalf\s^2 S^2
    \frac{\partial^2V(S,P,t)}{\partial S^2}\\
      &+& \sum_{r=1}^R \m_r \bigg(
  S\frac{\partial V(S,P,t)}{\partial S}
  -\frac{V(\a^r S,P,t)-V(S,P,t)}{\a^r-1}
  \bigg)
\end{eqnarray*}
In other words, the price of such a contract will at all times
remain a function of $S$ and $P$ only, and the PDDE reduces
to two dimensions. It is very important to notice that this
does not imply that the contract can be hedged using $S$ and $P$
only. In fact, the exact hedge ratios are given by
\[
  \phi_S = \frac{\partial V(S,P,t)}{\partial S}, \hspace{5mm}
  \phi_{P^r} = \frac{V(\a^rS,P,t)-V(S,P,t)}{P^r(\a^r-1)}
  -\frac{S}{P^r}\frac{\partial V(S,P,t)}{\partial S}
\]
and $\phi_P$ follows from the portfolio replication condition.
Since we can use only $S$ and $P$ in our hedge, we might
consider to use a standard delta hedge
$\smash{\phi_S = \frac{\partial V(S,P,t)}{\partial S}}$,
$\smash{\phi_P = \frac{\partial V(S,P,t)}{\partial P}}$.
In that case, we are effectively left with a position in $P$
and the virtual tradables $P^r$. So we should use the freedom
to choose the drifts $\m_r$ in such a way that we get a `satisfactory'
return on the residual risk. Obviously, this choice will depend
upon the real world intensities of the Poisson processes and
upon our idea of a satisfactory return. 

\vspace{1\baselineskip}\noindent
To connect our formalism to pricing in a market driven by a
Levy process, we take a limit where the number of Poisson
processes goes to infinity. This amounts to the replacement
of the sum over $r$ by an integral over jumpsizes $\a$, as follows
\begin{eqnarray*}
  \pda{V(S,P,t)}{t} &+& \shalf\s^2 S^2
    \frac{\partial^2V(S,P,t)}{\partial S^2}\\
      &+& \int_0^\infty \m(\a)\bigg(
  S\frac{\partial V(S,P,t)}{\partial S}
  -\frac{V\big(\a S,P,t\big)-V\big(S,P,t\big)}{\a-1}
  \bigg) d\a
\end{eqnarray*}
The resulting pricing equation is a partial integro differential
difference equation (PIDDE). Such equations are hard to solve
in general. One possible approach is to consider the integral as a
correction term to the pure diffusion equation, and make a series
expansion around $\a=1$. Note that there must be restrictions
on the choice of $\m(\a)$ in order for the equation to make sense.
Since these issues fall out of the scope of this article, we will
postpone detailed treatment to a later article.

\section{Restricting the number of jumps}
\label{sec:some-applications}

In the previous section we have considered markets driven
by both Wiener and Poisson processes. Of course, the latter
can make an infinite amount of jumps. For some price processes
this might be a reasonable assumption. But there are also
cases in which we might want to restrict the number of jumps.
For example, when modeling credit risk it is natural to
allow only one jump in the stochastic process that models this
risk. In this section we will consider the consequences of
such a restriction and how we can deal with them. We will
focus on a restriction to one jump, but note that it is
straightforward to generalize the discussion to any finite
number of jumps.

\subsection{Credit risk, Duffie-Singleton}

A fundamental difference between a market driven by a Poisson
process, which can make infinitely many jumps, and one driven
by a stopped Poisson process, which can make only one jump,
is that in the latter case the number of effective driving
processes is not constant in time: the stopped Poisson process 
will no longer be effective after the jump. Let us consider
a simple example, a market with a treasury bond $P^0$ and
a corporate bond $P^1$. Taking the treasury bond as numeraire
we assume the following dynamics {\it before} the jump
\begin{equation}
  \label{eq:28}
  \left. \begin{array}{rcl}
  \frac{dP^0_t}{P^0_{t_-}} & = & 0 \\
  \frac{dP^1_t}{P^1_{t_-}} & = & \m dt + (\a-1)\, dN_{t\wedge t_d}
  \end{array} \right\} \mbox{for $t\leq t_d$}
\end{equation}
where $N_{t\wedge t_d} \equiv N_{\min(t,t_d)}$ is a Poisson process
stopped after the first jump at time $t_d$, the time of
default. We see that at default, the corporate bond jumps like
\[
  P^1_{t_d} = \a P^1_{t_{d-}}
\]
i.e. the new value is a fraction of the old value.
This is called `recovery-of-market-value' by Duffie-Singleton
\cite{DuffieSingleton95}, and is a basic assumption
of their credit risk model. It is to be contrasted to
what they call `recovery-of-treasury', which is an assumption
of the Jarrow-Turnbull model \cite{JarrowTurnbull95},
to be considered in the next section. To avoid
arbitrage, we take the recovery rate $0<\a<1$ and the credit
spread $\m>0$. Now what happens after $t_d$? At that time we have
two tradables and no source of randomness. This represents
an overcomplete market, and we have to be careful not to
introduce arbitrage opportunities. For example, we cannot use
Eq.~(\ref{eq:28}) for $t>t_d$ because the ratio $P^1/P^0$ is
strictly increasing at that time. The most obvious way to deal
with this situation is to simply drop one of the tradables.
Indeed, after default, the corporate bond will no longer be
traded, so it effectively seizes to exist. Still, it will have
some value, and if the holder of the corporate bond does not
receive this value immediately at the time of default, it will
have to be reinvested in the remaining tradables. The most
natural thing to do is to reinvest the money in the treasury bond.
So we could say
\begin{equation}
  \label{eq:29}
  P^1_t = \frac{P^1_{t_d}}{P^0_{t_d}} P^0_t, \hspace{5mm} t>t_d
\end{equation}
Using this, we can formally extend the dynamics of the tradables
to all $t$, and find, in the same numeraire
\[
  \frac{dP^1_t}{P^1_{t_-}} =
  {\bf 1}_{t\leq t_d}(\m dt + (\a-1)\, dN_t)
\]
We now turn to the pricing problem in this market. It will
be obvious that the price of a contract will depend on the
information whether default has occurred or not. In fact, it
is easy to incorporate path-dependence in the problem, since
the path is fully specified by the time of default $t_d$.
Now before default, the price will depend on both $P^0$ and
$P^1$ and it will be written as $V_0(P^0,P^1,t)$. After default
it is useful to write it as $V_1(P^0,P^1,t_d,t)$, but we have
to keep the relation Eq.~(\ref{eq:29}) in mind. The derivation
of the pricing PDDE's follows the same lines as the general
discussion in the previous section and will not be repeated.
We find
\[
  \pda{V_0(P^0,P^1,t)}{t}+V_1\big(\b_0 P^0,\b_1 P^1,t,t\big) 
  -\sum_{i=0,1}\b_i P^i\pda{V_0(P^0,P^1,t)}{P^i}=0
\]
\[
  \pda{V_1(P^0,P^1,t_d,t)}{t}=0
\]
where
\[
  \b_0=\frac{\m}{1-\a}, \hspace{5mm}
  \b_1=\frac{\a\m}{1-\a}
\]
The hedge ratios before default are given by
\begin{eqnarray}
  \label{eq:30}
  \phi_0(P^0,P^1,t)&=&
  \frac{V_0(\b_1 P^0,\b_1 P^1,t)-V_1(\b_0 P^0,\b_1 P^1,t,t)}
  {P^0(\b_1-\b_0)} \\
  \label{eq:31}
  \phi_1(P^0,P^1,t)&=&
  \frac{V_1(\b_0 P^0,\b_1 P^1,t,t)-V_0(\b_0 P^0,\b_0 P^1,t)}
  {P^1(\b_1-\b_0)}
\end{eqnarray}
Let us now proceed to solve the pricing equations for the case
of a European security with maturity $T$. The payoff will be
specified by a function $V_0(P^0,P^1,T)$ for the case that no
default occurred before maturity and a function $V_1(P^0,P^1,t_d,T)$
in case default occurred at $t_d\leq T$. We now introduce
$\t\equiv T-t$, $\t_d\equiv T-t_d$, $y\equiv P^1/P^0$, 
$V_0(P^0,P^1,t)\equiv P^0 v_0(y,\t)$,
$V_1(P^0,P^1,t_d,t)\equiv P^0 v_1(y,\t_d,\t)$, and use homogeneity
to arrive at
\[
  \pda{v_0(y,\t)}{\t} +(\b_1-\b_0) y \pda{v_0(y,\t)}{y} 
  - \b_0 \bigg(v_1\bigg(\frac{\b_1}{\b_0} y,\t,\t\bigg)
  -v_0(y,\t)\bigg) = 0
\]
\[
  \pda{v_1(y,\t_d,\t)}{\t} = 0
\]
The second equation leads to
\[
  v_1(y,\t_d,\t) = v_1(y,\t_d,0)
\]
Inserting this in the first equation and making a change of
variables, introducing $z\equiv ye^{-(\b_1-\b_0)\t}$, we find
\[
  \pda{v_0(z,\t)}{\t} + \b_0 v_0(z,\t) =
  \b_0 v_1\bigg(\frac{\b_1}{\b_0}e^{(\b_1-\b_0)\t}z,\t,0\bigg)
\]
We now multiply by an integrating factor, then integrate
\[
  e^{\b_0\t}v_0(z,\t)=v_0(z,0)+\b_0\int_0^\t e^{\b_0s}
  v_1\bigg(\frac{\b_1}{\b_0}e^{(\b_1-\b_0)s}z,s,0\bigg)ds
\]
Rewriting this in terms of the original functions and
coordinates we get this symmetric result
\begin{eqnarray}
  V_0(P^0,P^1,t) &=& V_0(e^{-\b_0\t}P^0,e^{-\b_1\t}P^1,T) \nn \\
  \label{eq:32}
  &+&\int_t^T V_1(\b_0e^{-\b_0(s-t)}P^0,\b_1e^{-\b_1(s-t)}P^1,
  s,T)ds
\end{eqnarray}

\subsection{Credit risk, Jarrow-Turnbull}

Next, we consider a variation on the model treated above, now
using the assumption of `recovery-of-treasury'. As we mentioned
earlier, this is a basic assumption of the Jarrow-Turnbull
credit risk model \cite{JarrowTurnbull95}. We find that the
pricing equations have exactly the same form as in the previous
section, unifying the two approaches in one framework. Now
under `recovery-of-treasury' the corporate jumps, at the time of
default, like
\[
  P^1_{t_d} = \a P^0_{t_{d-}}
\]
i.e. the new value is a fraction of the value of the treasury
bond. This corresponds to the following choice of dynamics
\[
  \left. \begin{array}{rcl}
  \frac{dP^0_t}{P^0_{t_-}} & = & 0 \\
  \frac{dP^1_t}{P^1_{t_-}} & = & \m dt + 
  \bigg(\frac{\a P^0_{t_-}}{P^1_{t_-}}-1\bigg)\, dN_{t\wedge t_d}
  \end{array} \right\} \mbox{for $t\leq t_d$}
\]
So effectively we now have a tradable dependent jumpsize. We
can now literally follow the discussion for the Duffie-Singleton
case, and find that the pricing PDDE's and hedge ratios have exactly
the same form except that the $\bbe$ are now dependent on the
tradables as follows
\[
  \b_0=\frac{\m P^1}{P^1-\a P^0}, \hspace{5mm}
  \b_1=\frac{\a\m P^0}{P^1-\a P^0}
\]
Note that if we assume that $\m>0$, the model is arbitrage
free iff $P^1>\a P^0$. This relation cannot be true at all times.
In fact there will be a $t_c$ such that $P^1_{t_c}=\a P^0_{t_c}$ and
the model is arbitrage free only {\it after} this $t_c$. This is a
fundamental problem of the Jarrow-Turnbull approach. However,
for most realistic problems, time will be in the proper range
and we need not bother about the problem. Again we can solve the
pricing PDDE's for the case of a path-dependent European security.
Omitting the details of this calculation, we get
\begin{eqnarray}
  V_0(P^0,P^1,t) &=& 
  V_0\bigg(\frac{(\b_0-\b_1)P^0}{e^{(\b_0-\b_1)\t}\b_0-\b_1},
           \frac{(\b_1-\b_0)P^1}{e^{(\b_1-\b_0)\t}\b_1-\b_0},T\bigg)
  \nn \\ \label{eq:33}
  &+&\int_t^T \frac{(\b_0-\b_1)^2 e^{(s+t)(\b_0+\b_1)}}
  {(e^{\b_0s+\b_1t}\b_0-e^{\b_0t+\b_1s}\b_1)^2}V_1(\b_0 P^0,\b_1 P^1,s,T)ds
\end{eqnarray}
Note that if the payoff does not depend on the time of default,
so that we have $V_1(P^0,P^1,s,T)=V_1(P^0,P^1,T)$ for all $s$,
the integral can be evaluated explicitely, and we get
\begin{eqnarray*}
  V_0(P^0,P^1,t) &=& 
  V_0\bigg(\frac{(\b_0-\b_1)P^0}{e^{(\b_0-\b_1)\t}\b_0-\b_1},
           \frac{(\b_1-\b_0)P^1}{e^{(\b_1-\b_0)\t}\b_1-\b_0},T\bigg)\\
  &+& V_1\bigg(
  \frac{\b_0(e^{\b_0\t}-e^{\b_1\t})P^0}{\b_0e^{\b_0\t}-\b_1e^{\b_1\t}},
  \frac{\b_1(e^{\b_1\t}-e^{\b_0\t})P^1}{\b_1e^{\b_1\t}-\b_0e^{\b_0\t}},
  T\bigg)
\end{eqnarray*}
This simplification is what makes the Jarrow-Turnbull model
attractive for path-independent options. However, for path-dependent
options, the Duffie-Singleton model usually leads to more simple
results, as we will see in the next section.

\subsection{A credit default swap}
\label{sec:credit-default-swap}

Let us consider as an application a credit default swap (CDS)
\cite{Tavakoli98}. This contract can be viewed as a default insurance
on a corporate bond. The buyer of protection receives the difference
in value before and after default in case the issuer of the
corporate bond defaults during the lifetime of the CDS. In return,
he has to pay the seller of protection an annuity premium until the
time of default, or the maturity of the CDS, whichever comes first.
The contract can symbolically be decomposed in more elementary
contracts, each corresponding to a single cashflow, as follows
\[
  \mbox{CDS} = \mbox{Default} - \sum_i \a_i \mbox{Premium}(t_i)
\]
Here `$\mbox{Default}$' corresponds to the insurance payoff in
case of default, while the cashflow `$\mbox{Premium}(t_i)$'
corresponds to paying a premium of one unit of $P^0$ at time $t_i$
iff no default occurs before $t_i$. The $\a_i$ are usually
chosen in such a way that the initial value of the contract is
zero. Now let us assume that the treasury bond, the corporate
bond and the CDS all mature at the same time $T$, and (without
loss of generality) that the treasury and the corporate bond
have the same face value. This means that before default we have
the relation
\begin{equation}
  \label{eq:34}
  P^1 = e^{-\m(T-t)} P^0
\end{equation}
Using this, the payoff of `$\mbox{Default}$' can be described as
\begin{eqnarray*}
  V_0(P^0,P^1,T) &=& 0 \\
  V_1(P^0,P^1,t_d,T) &=& e^{-\m(T-t_d)}P^0-P^1
\end{eqnarray*}
The boundary conditions for `$\mbox{Premium}(t_i)$' are given by
\begin{eqnarray*}
  V_0(P^0,P^1,T) &=& P^0 \\
  V_1(P^0,P^1,t_d,T) &=& {\bf 1}_{t_i<t_d}P^0
\end{eqnarray*}

\subsubsection{Duffie-Singleton}

It is now straightforward to find the values of the
constituent parts of the CDS in the Duffie-Singleton model
before default by plugging the corresponding payoffs in
Eq.~(\ref{eq:32}), and simplifying the result using
Eq.~(\ref{eq:34}). We get
\begin{eqnarray}\label{eq:35}
\mbox{Default}(P^0,P^1,t) &=& 
(1-e^{-\b_1(T-t)})\frac{\b_0-\b_1}{\b_1}P^1 \\
\mbox{Premium}(t_i)(P^0,P^1,t) &=& \left\{ \begin{array}{ll}
e^{-\b_0(t_i-t)}P^0 & \mbox{if $t<t_i$} \\
P^0 & \mbox{otherwise} \end{array} \nn \right.
\end{eqnarray}
Hedge ratios follow from Eqs.~(\ref{eq:30},\ref{eq:31}), taking
$V_1(\b_0P^0,\b_1P^1,t,t)$ equal to $(\b_0-\b_1)P^1$ for the
`$\mbox{Default}$' contract and equal to ${\bf 1}_{t_i<t}\b_0P^0$
for `$\mbox{Premium}(t_i)$'.

\subsubsection{Jarrow-Turnbull}

And similarly for the Jarrow-Turnbull model, using
Eq.~(\ref{eq:33})
\begin{eqnarray*}
\mbox{Default}(P^0,P^1,t) &=& 
\ln\left(\frac{\b_1-\b_0}{\b_1-e^{-(\b_1-\b_0)(T-t)}\b_0}\right)
\frac{\b_1-\b_0}{\b_0}P^1 \\
\mbox{Premium}(t_i)(P^0,P^1,t) &=& \left\{ \begin{array}{ll}
\frac{\b_1-\b_0}{\b_1-e^{-(\b_1-\b_0)(t_i-t)}\b_0}P^0 &
\mbox{if $t<t_i$} \\ P^0 & \mbox{otherwise} \end{array} \right.
\end{eqnarray*}

\subsection{First-to-default insurance}
\label{sec:first-defa-insur}

In this section we consider the problem of pricing a first-to-default
insurance contract on a set of $R$ corporate bonds. This contract
pays the loss on the first bond to default. If none of the bonds
defaults during the lifetime of the contract it expires worthless
\cite{Tavakoli98}. We restrict our attention to corporate bonds under
the recovery-of-market-value convention, i.e. the Duffie-Singleton
approach. We assume the following dynamics taking the treasury
bond $P^0$ as numeraire (with $i=1,\ldots,r$)
\begin{equation}
  \label{eq:36}
  \frac{dP^i_t}{P^i_{t_-}} =
  \m_i dt + \rsum (\a^r_i-1)\, dN^r_{t\wedge t_d^r},
  \hspace{5mm} \mbox{for $t\leq \min(t_d^r)$}
\end{equation}
where $t_d^r$ is the time of default of $P^r$. A few remarks
are in place. It is assumed that the process $N^r$ models the
default process of corporate bond $P^r$, so that if $N^r$
jumps, $P^r$ seizes to exist. However, we do allow $N^r$ to
influence the prices of the other corporate bonds via the
elements $\a^r_i$ with $i\neq r$, thus introducing
a form of default correlation. Because of the simple form
of the contract, we will not need to know the dynamics of
the bonds after any default. Now the price function $V_0(\bx,t)$
before occurrence of any default should satisfy
\begin{equation}
  \label{eq:37}
  \pda{V_0(\bx,t)}{t}+\sum_{r=1}^R \bigg(V_r\big(\bbe^r \bx,t\big) 
  -\sum_{i=0}^N\b^r_i x_i\pda{V_0(\bx,t)}{x_i}\bigg) = 0
\end{equation}
where $V_r(\bx,t)$ is the price of the contract given that
corporate bond $P^r$ has defaulted. For a general contract
these functions could themselves be complex derivatives on
the remaining tradables. In the present case they are very
simple indeed
\[
  V_r(\bx,t) = \left(\frac{1}{\a^r}-1\right) P^r 
\]
Inserting this in Eq.~(\ref{eq:37}) we are left with
\[
  \pda{V_0(\bx,t)}{t}+\sum_{r=1}^R \bigg((\b^r_0-\b^r_r)P^r
  -\sum_{i=0}^N\b^r_i x_i\pda{V_0(\bx,t)}{x_i}\bigg) = 0
\]
To find the price of our contract, we integrate this PDE
using the boundary condition $V_0(\bx,T)=0$. The result is
\[
  V_0(\bx,t) = \rsum \frac{(1-e^{-\b_r\t})
  (\b^r_0-\b^r_r)P^r}{\b_r}, \hspace{5mm}
  \b_i=\rsum \b_i^r
\]
It is interesting to observe that every term in this sum
has the same functional form as a default insurance on a single
corporate bond, as can be seen by comparing with Eq.~(\ref{eq:35}).

\subsection{Consistent default correlation}
\label{sec:cons-defa-corr}

The first-to-default insurance contract was relatively easy to
price because of the special form of its payoff. We will now
consider the pricing problem for a more general type of contract.
The fundamental problem to be faced is how to formulate a
model for credit-risk that is self consistent, not only before
occurrence of any default, but at all times, in the presence of
default correlation. We start out by generalizing
Eq.~(\ref{eq:36}) as follows
\[
  \frac{dP^i_t}{P^i_{t_-}} =  \m_i({\bf t_d},t) dt
  + \rsum (\a^r_i({\bf t_d},t)-1)\, dN^r_{t\wedge t_d^r} 
\]
where ${\bf t_d}=\{t_d^1,\ldots,t_d^r\}$ is the set of default
times. For the sake of simplicity we assume that the credit
spreads and jump sizes only depend on time (and default information)
but not on other tradables. These functions should satisfy
some self-consistency conditions, which follow from the simple
observation that if a given corporate bond does not default during its
lifetime, it will have a definite value at maturity,
{\it independent} of possible default events of other bonds. 
So integration of the dynamic equations of a bond $P^i$ up to
maturity should yield the same value for any possible path for
which this bond does not default. Let us illustrate this by looking
at a simple example. We look at two corporate bonds $P^1$ and $P^2$
and a treasury bond $P^0$. Taking the latter as numeraire, we write
the dynamics as
\[
  \frac{dP^1_t}{P^1_{t_-}} = \left\{ \begin{array}{ll}
  \m_1(t)dt + \sum_{r=1,2} (\a_1^r(t)-1) dN^r_{t\wedge t_d^r} 
  & \mbox{for $t\leq\min(t_d^1,t_d^2)$} \\
  \m_1(t_d^2,t)dt + (\a_1^1(t_d^2,t)-1) dN^1_{t\wedge t_d^1} 
  & \mbox{for $t_d^2<t\leq t_d^1$} \\
  0 & \mbox{for $t_d^1<t$} \end{array} \right.
\]
\[
  \frac{dP^2_t}{P^2_{t_-}} = \left\{ \begin{array}{ll}
  \m_2(t)dt + \sum_{r=1,2} (\a_2^r(t)-1) dN^r_{t\wedge t_d^r} 
  & \mbox{for $t\leq\min(t_d^1,t_d^2)$} \\
  \m_2(t_d^1,t)dt + (\a_2^2(t_d^1,t)-1) dN^2_{t\wedge t_d^2} 
  & \mbox{for $t_d^1<t\leq t_d^2$} \\
  0 & \mbox{for $t_d^2<t$} \end{array} \right.
\]
Now suppose that $P^1$ matures at time $T$. The consistency
relations for this bond then take the form (assuming $t<t_d^2$)
\begin{eqnarray*}
  P^1(T) &=& P^1(t) \exp\left(\int_t^T \m_1(s) ds\right) \\
         &=& P^1(t) \a_1^2(t_d^2) 
  \exp\left(\int_t^{t_d^2} \m_1(s) ds
           +\int_{t_d^2}^T \m_1(t_d^2,s) ds\right)
\end{eqnarray*}
or equivalently
\[
  \a_1^2(t_d^2) = \exp\left(\int_{t_d^2}^T
  \big(\m_1(s)-\m_1(t_d^2,s)\big) ds\right)
\]
and similarly for $P^2$
\[
  \a_2^1(t_d^1) = \exp\left(\int_{t_d^1}^T 
  \big(\m_2(s)-\m_2(t_d^1,s)\big) ds\right)
\]
So we see that the off-diagonal jump sizes are completely fixed
by a choice of the drift functions, and also that if one bond
jumps down in response to default of another, its credit spread
must go up, just as one would expect. Now consider the pricing
problem. Again, the price of a contract will depend on information
which of the bonds has defaulted and when, and we need four
functions to describe it. The price $V_0(\bx,t)$ before any default
should satisfy
\[
  0 = \pda{V_0(\bx,t)}{t} + \sum_{r=1,2} \bigg(
  V_r\big(\bbe^r(t)\bx,t,t\big)
  -\sum_{i=0}^2\b^r_i(t) x_i\pda{V_0(\bx,t)}{x_i}\bigg)
\]
where the $\b^r_i(t)$ are found in the standard way from
the dynamics before any default. The functions
$V_r(\bx,t_d^r,t)$ with $r=1,2$ correspond to the price given
that only bond $P^r$ has defaulted at time $t_d^r$. They satisfy
\begin{eqnarray*}
  0 &=& \pda{V_1(\bx,t_d^1,t)}{t} + 
  V_{12}\big(\bbe^2(t_d^1,t)\bx,t_d^1,t,t\big)
  -\sum_{i=0}^2\b^2_i(t_d^1,t)x_i\pda{V_1(\bx,t_d^1,t)}{x_i} \\
  0 &=& \pda{V_2(\bx,t_d^2,t)}{t} + 
  V_{12}\big(\bbe^1(t_d^2,t)\bx,t,t_d^2,t\big)
  -\sum_{i=0}^2\b^1_i(t_d^2,t)x_i\pda{V_2(\bx,t_d^2,t)}{x_i}
\end{eqnarray*}
In this case, $\b^1_0(t_d^2,t)$ and $\b^1_1(t_d^2,t)$ are found
from the dynamics of $P^0$ and $P^1$ given that $P^2$ has defaulted
at time $t_d^2$ and $P^1$ has not. Next, $\b^1_2(t_d^2,t)$
should be taken equal to $\b^1_0(t_d^2,t)$, corresponding to
the fact that $P^2$ has the same dynamics as the treasury. A similar
story holds for the $\b^2_i(t_d^1,t)$. Finally, the price
$V_{12}(\bx,t_d^1,t_d^2,t)$ after default of both corporate bonds
simply satisfies
\[
  0 = \pda{V_{12}(\bx,t_d^1,t_d^2,t)}{t}
\]
In general, this set of equations will be hard to solve
and we will have to revert to numerical techniques. It is a challenge
to find model which is analytically managable, while still
incorporating default correlation in a realistic way. We will
return to this problem in future work.

\section{Conclusions and outlook}
\label{sec:conclusions-outlook}

Using local scale invariance, a.k.a. numeraire invariance, as a first
fundamental principle we have shown how one can derive derivative
security prices, and hedge ratios, in a complete market with
underlying price processes driven by Wiener and Poisson processes. We
discussed the various symmetries that should be satisfied by the
dynamic equation governing the prices of derivative securities and the
subtle differences regarding the notion of `market price of risk'
between the Wiener and Poisson case. We have further indicated how to
extend the jump-diffusion case to Levy processes. The complete market
case provides a natural basis to introduce the effects of incomplete
markets. We showed how this leads to PIDDE where a `market-price' of
risk has to be specified via some function on the jump-size. This
provides a natural and intuitive starting point for generating
corrections to the usual Wiener case. When applied to stock options,
such models give an alternative way to explain and model the volatility
surface, which seems to be in better agreement with the observed
behaviour then the standard `local volatility' approach 
\cite{AndersenAndreasen00}.

Another application that bears the fruit of making the symmetries
explicit is credit derivatives. In that case we were able to show
how both the Duffie-Singleton and Jarrow-Turnbull approaches can be
understood from one encompassing framework. Furthermore we introduced
a `market model' for credit risk and showed how to compute
first-to-default insurance contracts in such a framework. The
important notion of credit correlation can be modeled in a
consistent and straightforward manner using our framework.

Symmetries invoke constraints on model building and as such they
provide guidance in how to construct good models for derivative
security prices. The general pricing equation derived for Levy
processes provides a good starting point for generating solutions in
some perturbative expansion. Such approximations can be used to mark
the model to the market for example.

The credit correlation model can be computed numerically for arbitrary
choices of the parameters that determine the stochastic dynamics. It
is however important to have a simple, easily calculable model
that provides swift and accurate prices and hedge ratios for credit
derivative securities. 

We will come back to these points in future work.

\appendix

\section{Time dependent case}
\label{sec:time-dependent-case}

In this appendix we will give a proof by direct substitution
of the correctness of the general solution as given in
Sec.~\ref{sec:general-solution}. For a more direct derivation
we refer to the results obtained in Ref.~\cite{HooglandNeumann99a}
(for the diffusion part) and Sec.~\ref{sec:solving-pdde}
(for the jump part). The PDDE to be solved is given by
\begin{eqnarray*}
0 &=& 
  \pda{V(\bx,t)}{t} + \shalf
  \sum_{i,j=0}^N\sum_{m=1}^M \s_i^m(t)\s_j^m(t)x_ix_j\pdb{V(\bx,t)}{x_i}{x_j}
  \nn
  \\
  &&
  + \sum_{r=1}^R \bigg(
  V\big(\bbe^r(t) \bx,t\big) 
  -\sum_{i=0}^N\b^r_i(t) x_i\pda{V(\bx,t)}{x_i}\bigg).  
\end{eqnarray*}
where the $\b_i^r(t)$ are defined by
\[
  \b^r_i(t) \equiv \a_i^r\tl_r(t), \hspace{5mm}
  \tl_r(t) \equiv -\sum_{j=0}^N \m_j(t) A_{j,M+r}^{-1}(t) 
\]
Note that the $\a_i^r$ are assumed to be constant. We will see
that this assumption is crucial to get a simple form for the
solution. The reason is that if jumpsizes are not constant,
effects of jumps can no longer be described in terms of the
number of jumps that occur, but one needs exact information on
the timing of the jumps. To solve the PDDE, we use the following
ansatz for the solution
\begin{equation}
  \label{eq:38}
  V\big(\bx,t\big) = \nsum \int_{{\bf R}^M}
  g\big(\{x_i\,\phi(\bz-\bth_i)\, \pi_\bn(\brh_i)\}\big) d\bz
\end{equation}
Here $g(\bx)=V(\bx,T)$ describes the boundary condition at maturity
$T$. The vectors $\bth_i\equiv(\theta^1_i,\ldots,\theta^M_i)$ follow
from a singular value decomposition of the time-integrated covariance
matrix
\[
  \sum_{m=1}^M \theta_i^m(t)\theta_j^m(t)\equiv  
  \int_t^T \sum_{m=1}^M \s_i^m(u)\,\s_j^m(u)\,du
\]
In a similar way, the vectors
$\brh_i\equiv(\r^1_i,\ldots,\r^R_i)$ are defined as
\[
  \r_i^r(t) 
  \equiv \int_t^T \b^r_i(u)\,du
\]
The $\phi(\bz)$ and $\p_\bn(\brh_i)$ are defined by
\begin{eqnarray*}
  \phi(\bz)
  &\equiv&
  \prod_{m=1}^M \frac{1}{\sqrt{2\p}} e^{-\half z_m^2}
  \\
  \pi_\bn(\brh_i) 
  &\equiv& 
  \prod_{r=1}^R
  \frac{\big(\r_i^r\big)^{n_r}}{n_r!} \,
  e^{-\r_i^r}
\end{eqnarray*}
To prove that this ansatz is indeed the solution of the PDDE we
consider the two seperate cases, where the jumps and diffusions are
switched off respectively. Leibnitz' rule then suffices to prove that
the case with both diffusions and jumps is solved by the ansatz
Eq.~(\ref{eq:38}). We start with the diffusion part, showing that the
function
\[
  V\big(\bx,t\big) = \int_{\bR^M}
  g\big(\{x_i\,\phi(\bz-\bth_i)\}\big) d\bz
\]
solves the PDE
\[
  0=\cL V(\bx,t)=\pda{V(\bx,t)}{t}+\shalf
  \sum_{i,j=0}^N\sum_{m=1}^M \s_i^m(t)\s_j^m(t)x_ix_j\pdb{V(\bx,t)}{x_i}{x_j}
\]
To this end, we need the following identities,
which are easy to derive
\begin{eqnarray}
  \label{eq:39}
  \pda{}{t}\left(\sum_{m=1}^M \theta_i^m(t)\theta_j^m(t)\right) &=& 
  -\sum_{m=1}^M \s_i^m(t)\,\s_j^m(t) \\
  \pda{\phi(\bz)}{z_m}&=&-z_m\phi(\bz) \nn \\
  \label{eq:40}
  \pda{g}{t} &=& \sum_{i,m} (z_m-\theta_i^m(t))
  \pda{\theta_i^m(t)}{t} x_i \pda{g}{x_i} \\
  \label{eq:41}
  \pda{g}{z_m} &=& \sum_i (z_m-\theta_i^m(t)) x_i\pda{g}{x_i}
\end{eqnarray}
From these we can derive the crucial equation
\begin{eqnarray*}
  \cL g &=& \sum_{i,m} (z_m-\theta_i^m(t))
  \pda{\theta_i^m(t)}{t} x_i \pda{g}{x_i}
  -\sum_{i,j,m}\theta_i^m(t)\pda{\theta_j^m(t)}{t}
  x_ix_j\pdb{g}{x_i}{x_j}\\
  &=& \sum_{i,j,m} (z_m-\theta_i^m(t))\pda{\theta_j^m(t)}{t}
  \left(x_i\pda{}{x_i}\right)\left(x_j\pda{}{x_j}\right)g\\
  &=& \sum_{j,m} \pda{\theta_j^m(t)}{t}
  \left(x_j\pda{}{x_j}\right)\pda{g}{z_m}
\end{eqnarray*}
where Eqs.~(\ref{eq:39},\ref{eq:40}) were used in the first
step, homogeneity in the second and Eq.~(\ref{eq:41}) in the third.
If we plug this into the PDE we find
\[
  \cL V(\bx,t) = \int_{{\bf R}^M} \cL g d\bz =
  \sum_{j,m} \pda{\theta_j^m(t)}{t} \left(x_j\pda{}{x_j}\right)
  \int_{{\bf R}^M} \pda{g}{z_m} d\bz
\]
The integral over $z_m$ becomes trivial and we see that
the expression vanishes because of the strong damping of the
Gaussian at infinity.

\vspace{1\baselineskip}\noindent
Next, we consider the jump part, showing that the function
\[
  V\big(\bx,t\big) = \nsum 
  g\big(\{x_i\, \pi_\bn(\brh_i)\}\big) \equiv
  \nsum g_\bn
\]
solves the equation
\[
  0 = \cL V(\bx,t) =
  \pda{V(\bx,t)}{t}
  + \sum_{r=1}^R \bigg(
  V\big(\bbe^r(t) \bx,t\big) 
  -\sum_{i=0}^N\b^r_i(t) x_i\pda{V(\bx,t)}{x_i}\bigg)
\]
We first derive some useful identities
\begin{eqnarray}
  \label{eq:42}
  \pda{\p_\bn(\brh_i)}{t}
  &=& \sum_{r=1}^R \big( \p_\bn(\brh_i)-
  \p_{\bn-\d_r}(\brh_i){\bf 1}_{n_r\geq 1}
  \big) \b_i^r(t)
  \\
  \pi_n(\brh_i)\b_i^r(t) &=&
  \p_{\bn+\d_r}(\brh_i)\frac{\b_i^r(t)}{\r^r_i(t)}(n_r+1) \nn
\end{eqnarray}
where $\d_r$ is a vector with a one at position $r$ and zeros
elsewhere. The important thing to note in the last equation is
that the ratio of $\b_i^r$ and $\r_i^r$ does not depend on the
jumpsize $\a^r_i$ by virtue of the fact that the $\a^r_i$ are
constants, as can be easily seen from the definition of the $\b_i^r$'s:
\[
\frac{\b_i^r(t)}{\r^r_i(t)} = \frac{\tl^r(t)}{\int_t^T \tl^r(u)\,du}
\]
Therefore it does not depend on the index $i$ and this allows us to
exploit homogeneity again. Indeed, using the last equation and the
fact that $\pd_i g(\bx)$ is homogeneous of degree zero, making it
possible to scale out an arbitrary overall factor, we can write
\begin{eqnarray}
  \label{eq:43}
  \pd_i g\big(\{x_j\,\pi_\bn(\brh_j)\b_j^r(t)\}\big) 
  &=& \pd_i g_{\bn+\d_r}
\end{eqnarray}
where $\pd_i$ denotes the derivative w.r.t. the $i$-th argument. Since
$g(\bx)$ is homogeneous of degree $1$, $\pd_ig(\bx)$ is
homogeneous of degree $0$ and we can scale out an arbitrary overall
factor. By Eq.~(\ref{eq:42}) we find
\begin{eqnarray*}
  \pda{g_\bn}{t} &=& 
  \isum \pda{\pi_\bn(\brh_i)}{t} x_i \pd_i g_\bn \\
  &=&\rsum\isum \big( \p_\bn(\brh_i)-
  \p_{\bn-\d_r}(\brh_i){\bf 1}_{n_r\geq 1}
  \big) \b_i^r(t) x_i\pd_i g_\bn
\end{eqnarray*}
Homogeneity together with Eq.~(\ref{eq:43}) leads to
\begin{eqnarray*}
  \rsum g\big( \{ x_j\,\pi_\bn(\brh_j)\,\b_j^r(t) \} \big)
  &=& \rsum\isum
  \pi_\bn(\brh_i)\,\b^r_i(t) 
  x_i \pd_i g\big( \{ x_j\,\pi_\bn(\brh_j)\,\b_j^r(t) \} \big) \\
  &=& \rsum\isum \p_\bn(\brh_i) \b^r_i(t) x_i\pd_i g_{\bn+\d_r}
\end{eqnarray*}
Finally
\[
  -\rsum\isum \b^r_i(t) x_i\, \pda{g_\bn}{x_i}
  = -\rsum\isum \p_\bn(\brh_i) \b_i^r(t) x_i\pd_i g_\bn
\]
Putting this al together, we find after some renumbering
\[
  \cL V(\bx,t) = \nsum \cL g_\bn = 0
\]
and this concludes the proof.

\end{document}